\documentclass[12pt,preprint]{aastex}
\usepackage{amsmath}
\usepackage{rotating}
\usepackage{pdflscape}

\newcommand{\degree}{\ensuremath{^\circ}}

\begin{document}

\title{First Images of Cool Starspots on a Star Other than the Sun: Interferometric Imaging of $\lambda$ Andromedae}

\author{J. R. Parks\altaffilmark{1}, R. J. White\altaffilmark{1}, F. Baron\altaffilmark{1}, J. D. Monnier,\altaffilmark{2}, B. Kloppenborg\altaffilmark{1}, G. Henry\altaffilmark{3}, G. Schaefer\altaffilmark{1,4}, X. Che\altaffilmark{2}, E. Pedretti\altaffilmark{5}, N. Thureau\altaffilmark{5}, M. Zhao\altaffilmark{6}, T. ten Brummelaar\altaffilmark{4}, H. McAlister\altaffilmark{4}, S. T. Ridgway\altaffilmark{7}, N. Turner\altaffilmark{4}, J. Sturmann\altaffilmark{4}, L. Sturmann\altaffilmark{4}}
\altaffiltext{1}{Georgia State University, Department of Physics and Astronomy, 25 Park Place South, Suite 605, Atlanta, GA 30303-2911, USA}
\altaffiltext{2}{University of Michigan, Astronomy Department, 1805 S. University Ave., Ann Arbor, MI 48109-1090, USA}
\altaffiltext{3}{Tennessee State University, Center of Excellence in Information Systems, 3500 John A. Merritt Blvd., Box No. 9501, Nashville, TN 37203-3401, USA}
\altaffiltext{4}{The CHARA Array, Georgia State University, P.O. Box 3965, Atlanta, GA, 30302-3965, USA}
\altaffiltext{5}{University of St. Andrews, Department of Physics and Astronomy, Scotland, UK}
\altaffiltext{6}{Penn State University, Department of Astronomy and Astrophysics, 401 Davey Lab, University Park, PA 16802, USA}
\altaffiltext{7}{National Optical Astronomy Observatory, NOAO, Tucson, AZ, USA} 

\begin{abstract}
Presented are the first interferometric images of cool starspots on the chromospherically active giant $\lambda$ Andromedae.  These images represent the first model-independent images of cool starspots on a star other than the Sun to date.  The interferometric observations, taken with the Michigan Infra-Red Combiner coupled to the Center for High Angular Resolution Astronomy Array, span 26 days from Aug 17$^{th}$, 2008 to Sep 24$^{th}$, 2011.  The photometric time series acquired at Fairborn Observatory spanning Sep 20$^{th}$, 2008 to Jan 20$^{th}$, 2011 is also presented.  The angular diameter and power law limb-darkening coefficient of this star are 2.759 $\pm$ 0.050 mas and 0.229 $\pm$ 0.111, respectively.  Starspot properties are obtained from both modeled and SQUEEZE reconstructed images.  The images from 2010 through 2011 show anywhere from one to four starspots.  The measured properties of identical starspots identified in both the model and reconstructed images are within two $\sigma$ error bars in 51$\%$ of cases.  The cadence in the data for the 2010 and 2011 data sets are sufficient to measure a stellar rotation period based on apparent starspot motion.  This leads to estimates of the rotation period (P$_{2010}$ = 60 $\pm$ 13 days, P$_{2011}$ = 54.0 $\pm$ 7.6 days) that are consistent with the photometrically determined period of 54.8 days.  In addition, the inclination and position angle of the rotation axis is computed for both the 2010 and 2011 data sets; values ($\bar{\Psi}$ = 21.5$\degree$, $\bar{\emph{i}}$ = 78.0$\degree$) for each are nearly identical between the two years.
\end{abstract}

\keywords{stars:starspots \textemdash{} stars:imaging \textemdash{} stars:rotation \textemdash{} stars:individual($\lambda$ And) \textemdash{} methods:interferometric \textemdash{} methods:photometric{} \textemdash{} instrumentation:interferometers \textemdash{} techniques:image processing}

\section{INTRODUCTION}
\label{sec:intro}

In the decades since starspots were first hypothesized by \citet{kron47}, starspots have been studied, in detail, on scores of other stars.  More recently, space missions, such as \emph{Kepler}, have increased this number to potentially tens of thousands \citep{basri11}.  One thing that is certain is starspots are an ubiquitous phenomenon found on stars ranging in age from pre-main sequence to evolved giants \citep[and references therein]{strassmeier09}.  A motivation for studying starspots is a better understanding of stellar interiors, particularly the origins of magnetic dynamos.  Another motivation is that starspots complicate measurements of fundamental stellar properties (i.e. $T_{\mathrm{eff}}$, L, RV, etc.)  Besides the astrophysical importance, if a spotted star happens to harbor orbiting exoplanets, the increased uncertainties in the stellar properties will translate directly to increased uncertainties in the exoplanet properties (i.e. mass, radius).  With the advent of millimagnitude photometry, meter per second radial velocity surveys and direct milliarcsecond radius measurements, starspots as a ``second-order'' effect can no longer be ignored.

To account for the affects of starspots on particular measurements, the starspots themselves must be properly characterized.  The bulk of the current understanding of starspot properties (i.e. size, temperature, number, location) stems from two indirect observational techniques of magnetically active stars: light curve inversion and Doppler imaging.  Both methods suffer from certain assumptions that, if incorrect, can lead to unexpected starspot properties or imaging artifacts such as polar starspots and/or latitudinal starspot belts \citep{unruh97,berdyugina98}.  The assumptions include \emph{a priori} knowledge of the stellar inclination for accurate starspot latitude measurements, precise information of stellar parameters, accurate stellar atmosphere models, and accurate atomic and molecular line lists.  While the concern that these are artifacts rather than real features  has been largely addressed \citep{unruh96,rice02}, a more direct method for imaging starspots would bolster confidence in the present results.  

Laying aside the veracity of these techniques, the starspot characteristics these techniques have provided in terms of lifetimes, sizes, affect on stellar luminosity, and location is in many cases contrary to the behavior of sunspots.  For instance, large starspots are known to persist from months to years \citep{berdyugina05}.  However, typical sunspot lifetimes range from days to weeks.  The covering factor, or percentage of the visible surface covered by spots, is far larger for active stars (10$\%$ to 50$\%$) than for the Sun where the covering factor never exceeds 0.2$\%$ \citep{cox00}.  In addition, at times where the covering factor is largest, the overall luminosity of active stars decreases substantially ($\Delta$\emph{V} $\leq$ 0.6) whereas the Sun's overall luminosity actually increases.  Sunspots appear at a latitude of $\sim$30$\degree$ symmetric about the equator at the beginning of the solar activity cycle.  As the cycle progresses, sunspots migrate toward the equator stopping at a latitude of $\sim$8$\degree$ \citep[and references therein]{babcock61}.  Starspots have been observed to reside anywhere from low to high latitudes or at the poles \citep[and references therein]{strassmeier09}.  

Direct imaging of starspots is needed in order to confirm this discrepancy and potentially link the behavior of magnetic spots from solar-type stars to more active stars.  This direct measure of starspot properties can be obtained via long baseline optical/near infra-red interferometry (LBI).  By combining the light, akin to Young's double slit experiment, from multiple, widely spaced telescopes, sub-milliarcsecond angular resolutions can be achieved.  Since the first near-IR aperture synthesis image of the binary Capella by \citet{baldwin96}, images of binary stars, rapidly rotating stars, and star$+$disk systems are growing more commonplace \citep{tuthill01,hummel04,kloppenborg10,che11,baron12}.  Additionally, bright, convection-induced starspots have been imaged on the late type stars Betelgeuse and VX Sagittarii using LBI \citep{young00,chiavassa10}.  More recently similar starspots have also been imaged on the surfaces of both RS Per and T Per \citep{baron14}.  At present, the angular resolution of the longest baseline interferometer is $\sim$0.4 milliarcsecond (mas) in the H band.  This provides the highest imaging resolution by any instrument to date.  The median angular diameter for a survey of A, F and G main sequence stars is 0.991 mas or $\sim$2.5 resolution elements \citep{boyajian12}.  Therefore the technique is currently only viable for giant stars and close early-type dwarfs. 

\section{OBSERVATIONS AND DATA REDUCTION}
\label{sec:obs}

\subsection{The Chromospherically Active Giant $\lambda$ Andromedae}
\label{subsec:lam_and}

$\lambda$ Andromedae ($\lambda$ And; HD 222107) is a bright (\emph{V}: 3.872 mag, \emph{H}: 1.501 mag) G8 IV-III classified as a RS CVn type variable star in the Third Catalog of Chromospherically Active Binaries \citep{eker08}.  \citet{calder35} first discovered the photometric variability of $\lambda$ And with a historical peak $\Delta$\emph{V} amplitude of $\sim$0.3 magnitudes.  \citet{henry95} conducted a 15 year photometric monitoring campaign finding a periodic variability of 53.95 $\pm$ 0.72 days over an 11.1 $\pm$ 0.4 year stellar activity cycle.  $\lambda$ And was found by \citet{walker44} to be a single spectroscopic binary with an orbital period of 20.5212 days.  Mass estimates for $\lambda$ And in \citet[hereafter D95]{donati95} range from 0.65-0.85 M$_{\sun}$, and with considerable uncertainty.  From the mass function provided by the spectroscopic orbit, D95 infer a mass ratio \emph{q} = 0.12$^{+0.07}_{-0.04}$.  This in turns yields a companion mass of 0.08-0.10 M$_{\sun}$ leading to the conclusion the companion is a low main-sequence dwarf or high mass brown dwarf.  The high flux contrast between the two components of $\lambda$ And will preclude the companion affecting the photometric or interferometric observations.  \citet{nordgren99} measured a limb darkened angular diameter of $\lambda$ And of 2.66 $\pm$ 0.08 mas using the Naval Prototype Optical interferometer in the optical; the measurements spanned 10 spectral channels ranging from 649 to 849 nm.  This angular diameter is a factor of $\sim$5x larger than the \emph{H}-band angular resolution provided by the CHARA interferometer.  In short, $\lambda$ And is an interferometrically single, large, bright star with significant variability strongly believed to arise from cool starspots.

Doppler imaging of $\lambda$ And has not been possible due to low projected rotational velocity (\emph{v}sin(\emph{i}) = 6.5 km/s) insufficiently broadening absorption lines necessary to detect the deformations caused by starspots \citep{strassmeier09}.  On the other hand, light curve inversion has been modestly successful in studying starspots on $\lambda$ And.  \citet[hereafter F08]{frasca08} created a surface map via light curve inversion of optical photometry coupled with spectral line ratios.  This map shows two starspots, each covering $\sim8\%$ of the visible surface with temperatures $\sim$880 K cooler than the photosphere, separated by 81$\degree$ in longitude.  In addition, each starspot is preceded by a bright active region that is of comparable size to the starspot.

\subsection{Interferometric Observations}
\label{subsec:int_obs}

For those uninitiated with the terms and physics behind optical/near infra-red interferometry, an excellent review was written by \citet{monnier03}.  

$\lambda$ And was observed on 26 nights between Aug 17$^{th}$, 2008 and Sep 24$^{th}$, 2011.  Table~\ref{tab:obs_log} lists the date of the observations, the baselines utilized, the number of [u,v] points, and the calibrators (defined below) used on each night. The parenthetical number beside a calibrator indicates the number of times it was observed during the night.  Table~\ref{tab:cal_diam} contains the uniform disk angular diameters with error for each calibrator.  All observations were conducted using the Center for High Angular Resolution Astronomy (CHARA) array owned and operated by Georgia State University.  The array is composed of six 1-m telescopes in a non-redundant ``Y''-shaped configuration.  The baseline lengths range from 34 to 331 m, currently making this the longest baseline optical/near-infrared interferometer in the world \citep{tenbrummelaar05}; the longest baselines provide an angular resolution of $\sim$0.4 mas in the \emph{H}-Band.  The data were collected using the image-plane Michigan Infra-Red Combiner (MIRC) in the \emph{H}-band; (see \citet{monnier04,monnier06} for details).  A low resolution (R $\sim$ 42) prism splits the light into eight spectral channels with absolute wavelength precision of $\pm$0.25$\%$ based on measurements of $\iota$ Peg using the orbit of \citet{konacki10}.

Interferometric data are collected when the light path difference between each telescope pair is well below the coherence length.  This is achieved by ``delaying'' the light from one telescope by adding more path length to the light of the other telescope through the use of delay lines.  When the light paths are equal, the light from each telescope combines as an interference fringe.  The MIRC combiner is then set to track these fringes while the data frames are taken.  Afterwards a series of calibration frames are recorded that include a background and foreground frame, along with images of the light from each beam individually. Collection of the data and calibration frames typically does not exceed 30 minutes.  The total amount of data taken on a particular night is identical to the number of [u,v] points listed in Table~\ref{tab:obs_log}.  

The standard MIRC pipeline was used for data reduction \citep{monnier07}.  The frames containing the fringe pattern in each block of data were co-added.  These co-added frames are corrected for instrumental effects through a background frame subtraction and foreground frame normalization.  Raw squared visibilities, triple amplitudes, and closure phases are extracted using the Fourier transform of these corrected, co-added frames.  Photometric calibration due to differences in the flux amplitude per telescope beam is performed via real-time flux estimates derived from choppers that temporally encode the light from each telescope (prior to 2010) or through the use of a beam splitter following spatial filtering to shunt part of the beam to a CCD that directly monitors the flux levels from each telescope \citep[after 2010]{che10}.  The data are then transformed from relative measurements to absolute measurements through observations of a calibration star or ``calibrator''.  A calibrator is a star of known size that is typically on the order or smaller than the array's resolution limit and is within a few degrees to the target on the sky.

During the 2008 and 2009, only one or two snapshot observations were obtained of $\lambda$ And.  A snapshot observation is where only a single set of data frames bracketed by calibrator observations is taken.  While these observations were valuable confirming the presence of starspots from asymmetric closure phases, the starspots in the resultant images that cannot be confirmed as genuine.  The discussion of the observational strategy and results of these data sets can be found in Appendix~\ref{app:early_data}.

The 2010 observing run combined observations from the S1-E1-W1-W2 and S2-E2-W1-W2 telescope configurations.  Each telescope has an alphanumeric designation to describe its location.  The letter refers to the baseline direction (``S'' for south, ``E'' for east, ``W'' for west) while the number indicates location along that baseline, where ``1'' is exterior to ``2''.  The longest CHARA baseline is between E1 and W1.  The switch between telescope configurations occurred approximately at the mid-point of the night.  In both configurations $\lambda$ And was observed continuously with each observation bracketed by an observation of a calibrator.  This strategy of telescope configurations, constant monitoring and bracketed observations is designed to provide well calibrated measurements sampling the largest [u,v] coverage available.  This strategy yields 11 visibilities, 8 closure phases, and 8 triple amplitudes per spectral channel.

In addition to the above strategy, 2010 data sets consist of observations on sequential nights that were subsequently combined into a single OIFITS file.\footnote{OIFITS is the standard file format for optical/near-IR interferometric measurements \citep{pauls05}.}.  This strategy provides both an increased [u,v] coverage and a sanity check for the imaging methods since images of the star should be nearly identical between subsequent nights.  The only exception to this strategy is for Sept 10$^{th}$ and 11$^{th}$, as poor weather prevented observations on the 11$^{th}$.

The 2011 observing run benefited from the MIRC upgrade that enabled simultaneous 6 telescope observations.  Again to maximize the [u,v] coverage, $\lambda$ And was observed continuously over the night for as long as the delay lines would permit (typically 6 hrs for all 6 telescopes).  Again $\lambda$ And measurements were bracketed by measurements of a calibrator star.  Each individual observation yields 15 visibilities, 20 closure phases, and 20 triple amplitudes per spectral channel.  Since observations from subsequent nights were not combined, the number of [u,v] points obtained is approximately half of that of the 2010 data set, despite the addition of two telescopes.

\subsubsection{Interferometric Measurement Errors}
\label{subsubsec:int_err}

Standard errors are propagated for the measured quantities.  However, because of systematic effects two types of error are applied to the calibrated squared visibilities and triple amplitudes to appropriately account for systematics.   Additive errors are necessary for two different behaviors in the data.  The first is when the calibrated squared visibility and/or triple amplitude falls below zero.  As this is non-physical, a constant is added to enlarge the error to include zero.  The second is when the squared visibilities and triple amplitudes do not monotonically increase or decrease as a function of wavelength across the eight spectral channels.  The errors are then enlarged by a multiplicative constant to account for any abnormal structures (e.g. step-functions) found in the data across the 8 spectral channels.  Typical additive errors for the squared visibility and triple amplitude are $2 \times 10^{-4}$ and $1 \times 10^{-5}$, respectively.  The multiplicative errors improved after 2010 due to better photometric calibration provided by the photometric channels.  The typical multiplicative errors in squared visibility and triple amplitude are 15$\%$ (10$\%$) and 20$\%$ (15$\%$), respectively prior to 2010 (2011).

Typically 1$\degree$ is added to the closure phase errors, as suggested by \citet{zhao11}.  In addition, two additional closure phase errors are incorporated to avoid poor model fits due to calibration systematics.  These new errors are important in the low signal-to-noise (S/N) regime near to visibility null crossings.  As correlated camera readout noise dominates the closure phase measurements at low S/N, minimum closure phase errors are applied when the S/N in the triple amplitude signal is $\lesssim$ 1.  Finite time-averaging and spectral bandpass effects are accounted for by an error term proportional to $\Delta$CP$_{\lambda}$ across each spectral channel.  CP$_{\lambda}$ corresponds to the closure phase as a function of wavelength.  These two errors are applied to the closure phase noise via the following equation $\sigma_{CP} > MAX((30\degree/(S/N_{T3amp}^{2})$,0.2$\Delta$CP$_{\lambda}))$, where S/N$_{T3amp}$ is the signal-to-noise in the triple amplitude measurement.  

\subsubsection{Identification and Removal of Poor Data}
\label{subsubsec:poor_data}

In a few data sets in 2010 and 2011, the visibility in certain blocks of data on certain baseline pairs are not consistent with a simple Bessel function.  This inconsistency is not due to starspots as these features will only change the amplitude of the visibility lobes while keeping the form of the function intact.  The reason for these discrepant data blocks is most likely due to poor calibration on that baseline.  These discrepancies are found almost exclusively on the short baseline pairs (e.g. S1-S2, W1-W2, E1-E2).  The visibilities were either lower than the visibilities in the other blocks on the same baseline or possessed an opposite or flat slope in comparison to other blocks on the same baseline.  Data was judged to be discrepant via visual inspection and removed prior to either modeling or image reconstruction.  At most, the rejected data only amounted to 1$\%$ of the total data for any epoch. 

\subsubsection{37 And}
\label{subsubsec:37_and}

The star 37 And was used as a calibrator, in addition to other calibrators during the same night, for the 2008, 2009 and 2010 observing runs.  \cite{che12} discovered that 37 And is a high contrast binary based on sinusoidal-like variations in the closure phase with a few degree amplitude.

The orbit of the 37 And companion has been fully characterized by Roettenbacher et al. (2015, in preparation).  To study the effect this companion has on $\lambda$ And data, the visibility and closure phase of a single star (the assumed calibrator) and a binary system (the actual calibrator) are compared to each other.  The single star is taken to be a uniform disk with an angular diameter of 0.789 mas.  The binary system is composed of the single star and a point source with a flux ratio of 80 (primary/secondary) and a separation of 46.66 mas, the maximum component separation.  It is very likely the separation was shorter during the observations in question, however the full separation is used to determine the large possible error introduced.  The [u,v] coverage for this analysis is the same as Sep 2$^{nd}+$3$^{rd}$, which represents the densest coverage of any single epoch.

\begin{figure}
  \plotone{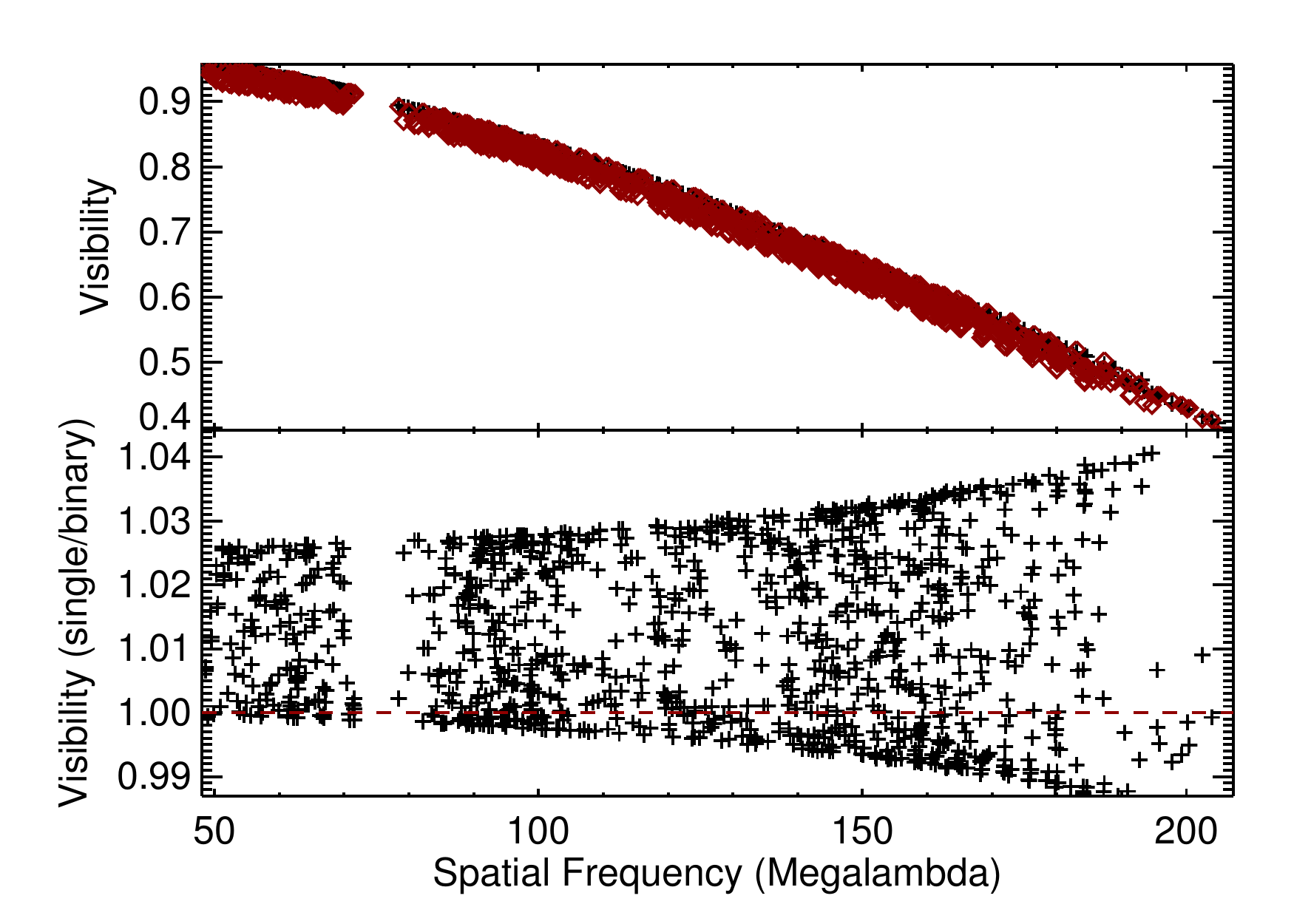}
  \caption{\emph{Top}: Shown here are both the visibilities expected from a uniform disk (0.798 mas) and the binary (point source secondary, flux contrast = 80, separation = 46.66 mas).  \emph{bottom}: This contains the uniform disk visibility divided by the binary visibility as a function of baseline.}
  \label{fig:37and_vis}
\end{figure}

Fig~\ref{fig:37and_vis} shows both the visibilities expected from the uniform disk and the binary.  Also shown is the uniform disk visibility divided by the binary visibility as a function of baseline.  The error introduced increases with baseline to a maximum of 4$\%$ with an average of $\sim$1.5$\%$  Given the typical multiplicative errors in squared visibilities are 10-15$\%$, this effect is considered negligible.  It should also be noted that this represents the worst contribution made by the binary and does not address the contribution of additional proper calibrators that will mitigate the above effect.

\begin{figure}
  \plotone{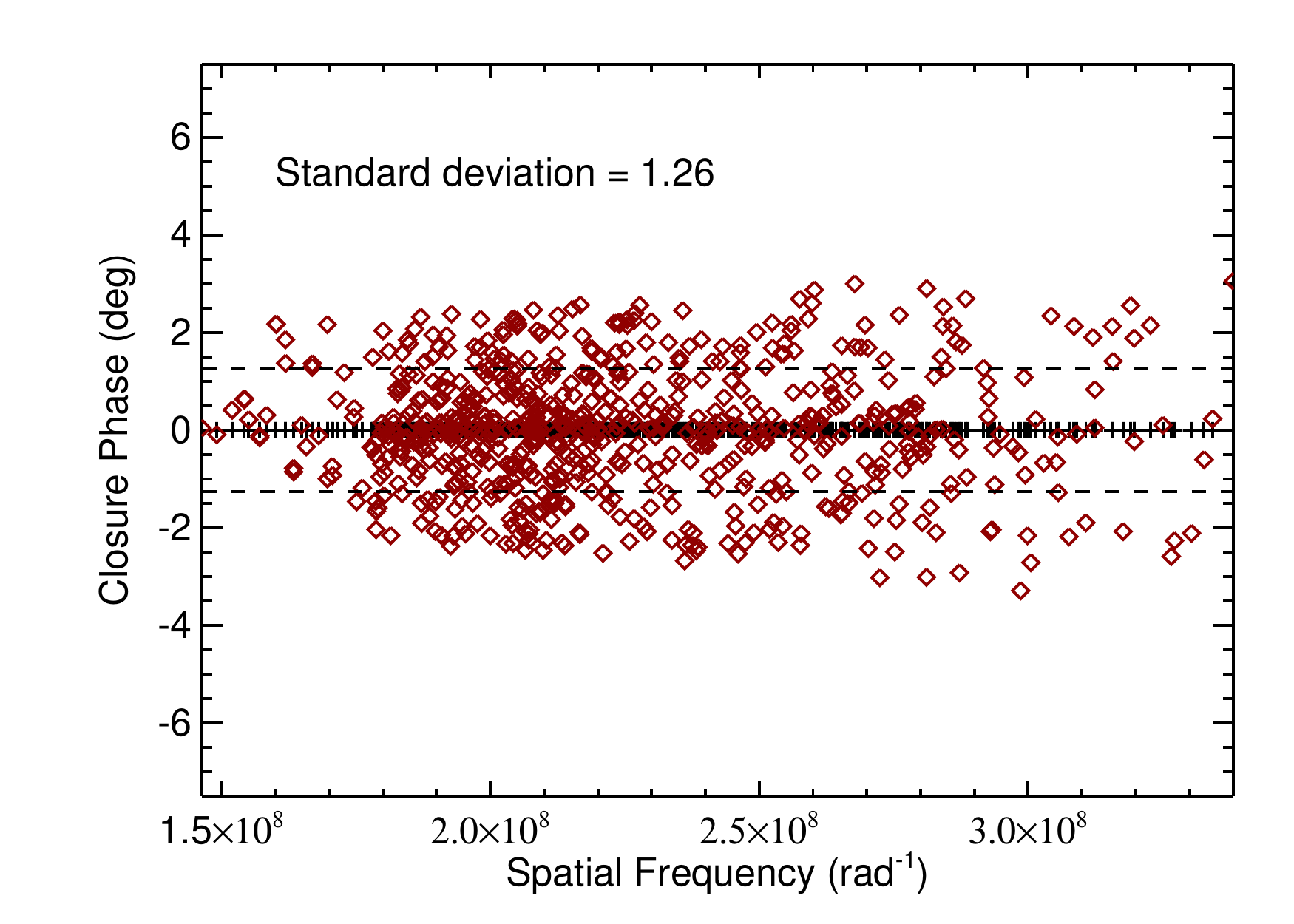}
  \caption{Shown here are both the closure phases expected from a uniform disk and the binary.  As expected the closure phases for the uniform disk are zero, while the standard deviation of the binary closure phases is 1.27$\degree$.}
  \label{fig:37and_cp}
\end{figure}

Fig~\ref{fig:37and_cp} shows the expected zero closure phase for the single uniform disk.  The red diamonds indicate the closure phase signal produced by the binary.  The standard deviation of this signal is 1.27$\degree$.  Given the noise floor for $\lambda$ And measured closure phase is 1$\degree$, the effect of the binary is negligible.

\subsection{Photometric Observations}
\label{subsec:phot_obs}

\begin{figure}
  \plotone{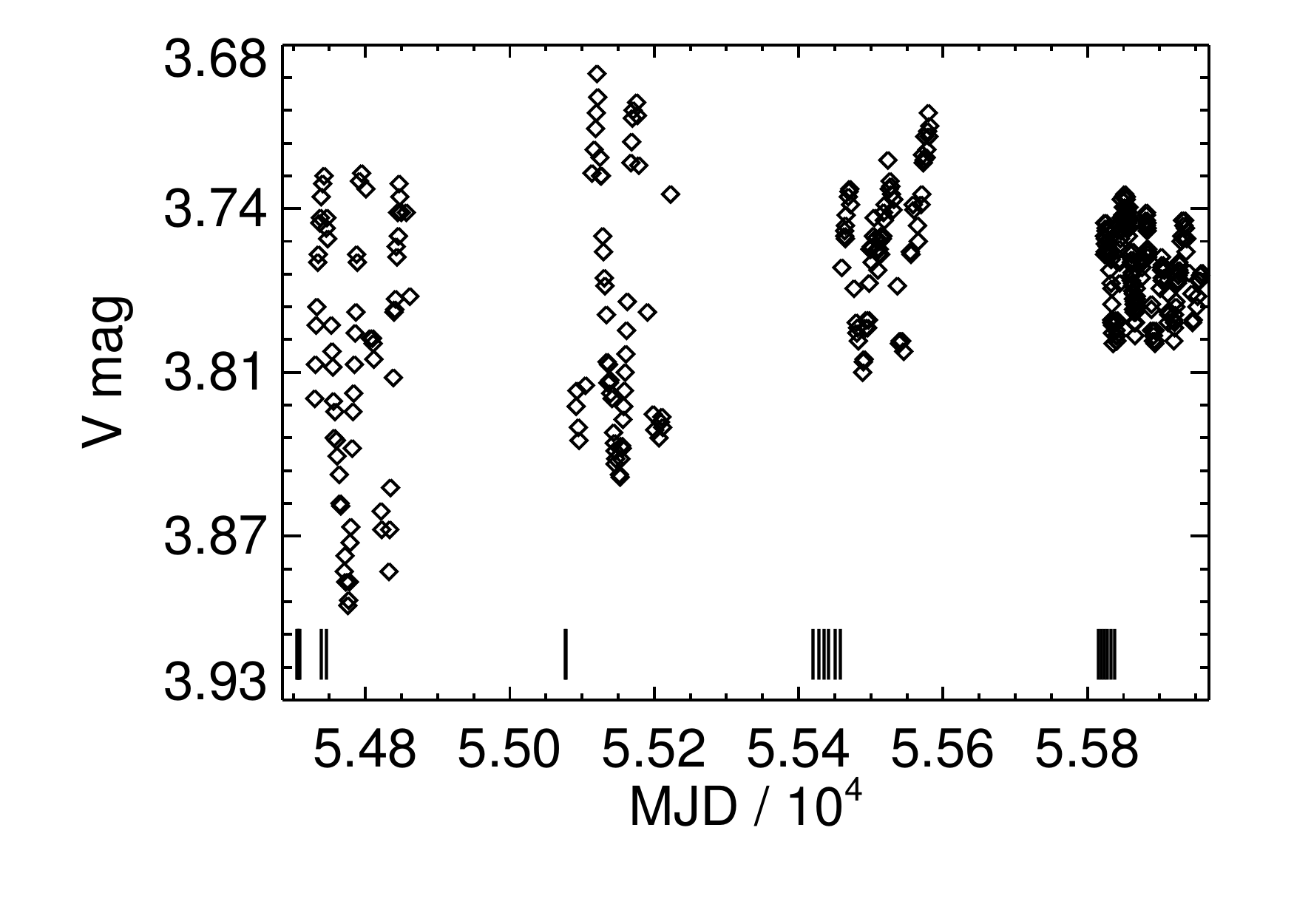}
  \caption{Time series photometry for $\lambda$ And ranging from Sep 2008 to Feb 2012.  The tick marks along the bottom axis represent the interferometric observations.  The brightening trend of the time-series is due to the 11.1 yr stellar cycle \citep{henry95}}
  \label{fig:photall}
\end{figure}

Photometric observations of $\lambda$ And were obtained using the 0.4-m automated telescope at Fairborn Observatory operated by Tennessee State University.  $\lambda$ And was observed 376 times for 3.4 years ranging from Sep 20$^{th}$, 2008 to Feb 1$^{st}$,  2012.  Fig~\ref{fig:photall} shows the measured time series.  The differential time series was corrected for atmospheric extinction and transformed into the Johnson \emph{V} filter system.  $\Psi$ And (\emph{V}: 4.982) and $\kappa$ And (\emph{V}: 4.137) were used as the comparison and  check star, respectively.  The typical photometric error is 6.0 millimag.  These errors are computed from the standard deviations in the time series of the check star.

\subsubsection{$\lambda$ And Optical Light Curve}
\label{subsubsec:light_curve}

The qualitative behavior of the time series (i.e. overall trends, changing variability amplitudes) is consistent with $\lambda$ And's 11.1 yr stellar cycle \citep{henry95}.  The time series can be split into 4 ``seasons''; each season corresponds to photometric observations taken just after the interferometric observations of the same year.  The offset between the beginning of the interferometric and photometric observations is $\sim$1 $\lambda$ And rotation.  The seasons are labeled for the year in which the observations were taken.  The time series period and variability amplitude in each season is measured for comparison with the modeled light curves generated from the interferometric images.  Table~\ref{tab:phot_log} contains the date ranges of each season, along with the identified period and the peak-to-trough $\Delta$\emph{V} amplitude.

The period of photometric variability for the time series in each season is identified using the Plavchan-Parks algorithm \citep{parks14}.  Fig~\ref{fig:photper} displays the light curve for each season folded to the most significant period.  Uncertainties in the period are set by the width of a Gaussian fit to the most significant peak in the power spectrum.  The four determined periods are 54.27 $\pm$ 0.032, 55.15 $\pm$ 0.91, 53.3 $\pm$ 1.1, and 53.3 $\pm$ 1.9 days.  The doubled appearance in season 4 could arise from starspots primarily on longitudes separated by $\sim$180$\degree$.  In the previous seasons, one or more starspots only dominate a single hemisphere.  The average rotation period of $\lambda$ And is 54.02 $\pm$ 0.88 days, where the reported error is the standard deviation of the four measured periods.

\section{STARSPOT ANALYSIS}
\label{sec:spot_anal}

Two different techniques are employed to characterize starspots from the observed interferometric data: a cool spotted stellar surface model and image reconstruction.  These methods are independent of each other; the results of one technique were \emph{not} used as a starting condition for the other. 

\subsection{Spotted Star Model}
\label{subsec:spot_model}

The stellar surface model is computed using an IDL code capable of modeling any number of circular cool or hot starspots on a limb-darkened surface.  In addition, the code accounts for the effects of fore-shortening on starspots located away from the sub-stellar point.  The free parameters are the stellar angular diameter ($\theta$), power law limb-darkening coefficient ($\alpha$), starspot covering factor ($\phi$), starspot latitude (\emph{b}), starspot longitude (\emph{l}), and starspot intensity ratio (\emph{f}).   The code extracts model interferometric data by computing the Fourier transform of an artificially generated stellar surface projected on the plane of the sky.  The sampling for the Fourier transform is taken from the [u,v] coverage of the observed data being modeled.  The goodness-of-fit parameter is the equally weighted reduced $\chi^{2}$ average between the observed and modeled visibilities, closure phases and triple amplitudes.

Changes in the angular diameter and, to a lesser extent, the limb-darkening coefficient have a large effect on modeled visibilities at spatial scales smaller than the first zero of the visibility function.  As the starspot information is contained on these small spatial scales, accurately determining the stellar properties prior to searching for the starspot properties is needed.  This is done by first combining all the interferometric data from 2010 and 2011 into a single OIFITS file.  The $\theta$ and $\alpha$ are measured by modeling only the visibility data using a grid search with a parameter step size of $1\times 10^{-3}$.  The parameter uncertainty is determined by the lowest reduced $\chi^{2}$ + 1 ellipse in the reduced $\chi^{2}$ space.  The furthest points of the ellipse along each axis are taken as the error in $\theta$ and $\alpha$, respectively.  Closure phases are not considered at this stage; $\lambda$ And is likely not rotationally (\emph{v}sin\emph{i} = 6.5 km/s) or Roche lobe distorted.

Once $\theta$ and $\alpha$ are known, model solutions for each epoch are computed using a multi-parameter downhill simplex minimization \citep[AMOEBA]{press92}.  Model solutions with one, two, and three starspots are run with the preferred model yielding the lowest $\chi^{2}$ statistic.  A fourth starspot model is only investigated if the presence of the additional starspot can be inferred from with the prior and subsequent epochs based on rotational grounds.  Only cool starspots are modeled as these are the type to persist on time-scales of a stellar rotation.  Early attempts with the AMOEBA algorithm on starspot models demonstrated that the solutions are biased by initial parameters and search scales.  This is indicative of numerous local minima in the reduced $\chi^{2}$ space along with the deeper global minimum.  The search scale employed is roughly 10$\%$ of the physical range for each parameter.  For example, the range in allowable intensity ratios is from 0.5 to 1.0 so the search scale is set to 0.05.  AMOEBA is only very effective at finding an accurate solution once the search occurs in the global minimum.  Therefore, a genetic algorithm is employed prior to running AMOEBA in order to find the approximate location of the reduced $\chi^{2}$ minimum.

A genetic algorithm (GA) is an iterative process through which a best solution is found by ``evolving'' an initial set of randomly chosen model solutions (members) \citep{charbonneau95}.  The fitness, or chance it will be used in the subsequent iteration, of each member is determined based on the member's reduced $\chi^{2}$.  The ``survival'' process is determined via a roulette wheel scheme.  The wheel is spun a number of times equal to the population size.  The probability the wheel will choose a member to survive is proportional to the member's fitness.  Therefore the next population will, in theory, be composed of model solutions with lower $\chi^{2}$ on average.  This new population is ``evolved'' via two different random methods: crossover and mutation.  Crossover takes sections of a parameter value and swaps it with another parameter value.  For example, solution A has a latitude of 45.12$\degree$ and solution B has a longitude of 12.57$\degree$.  Crossover can swap the digits after the decimal to yield a new latitude of 45.57$\degree$ and longitude of 12.12$\degree$.  Mutation causes a section of the parameter value to change randomly.  Using the previous example, the latitude 42.12$\degree$ could mutate to become 49.12$\degree$.  Both crossover and mutation are applied with a frequency of 90$\%$ and 1$\%$, respectively.  The fitness of the new population is determined and the entire process is iterated until the average fitness drops below a convergence criterion.

The reason AMOEBA is still needed after the GA is the GA's inability to converge to the exact minima in $\chi^{2}$ space.  During the GA and AMOEBA implementations, both the stellar angular diameter and limb-darkening coefficient are kept fixed.  

Errors to model starspot parameters are found by randomly varying each parameter of the best-fit model and then re-running the AMOEBA search algorithm.  All the parameters are uniformly randomized simultaneously within a given range.  $\phi$ and \emph{f} are allowed to vary by at most $\pm$0.1 while \emph{b} and \emph{l} are varied by 1.8$\degree$.  This procedure, which is meant to explore the reduced $\chi_{2}$-space around the best-fit solution, is performed ten times.  The parameter errors correspond to the standard deviation of the ten values yielded by the above procedure.  If a trial solution has a $\chi^{2}$ better than the initial final solution, this trial solution becomes the final solution.

Since the observations occur in the Rayleigh-Jeans tail of the spectral energy distribution, the temperature ratio between a starspot and the photosphere should be a linear function of the corresponding intensity ratio.  Therefore, the intensity ratio and its error are directly translated into the starspot temperature ratio, T$_{R}$, and corresponding error.

\subsection{SQUEEZE: Image Reconstruction}
\label{subsec:squeeze}

Model-based imaging is a very effective tool in determining starspot properties; however it is limited by the assumptions used to create the model (e.g. circular starspots).  Model-independent imaging, or image reconstruction, on the other hand has freedom to portray more realistic starspot shapes and sizes.  The main hurdle faced by image reconstruction is an ill-posed inverse problem brought on by working in a data starved regime, incomplete [u,v] coverage, and atmosphere-corrupted Fourier phase information.  Image reconstruction algorithms are all based on a regularized maximum likelihood paradigm that reconciles a $\chi^{2}$ statistic with \emph{k} number of regularization statistics, or regularizer ($R_{\mathrm{k}}$), modulated by a user-defined weighting parameter ($\mu_{\mathrm{k}}$).

\begin{equation}
    \label{eq:like_funct}
    \hat{i} = \underset{i\in\mathbb{R}^n}{\operatorname{argmax}}\Bigg\{\chi^2(i)+\sum_{k=1}^{K}\mu_{k}R_{k}(i)\Bigg\}
\end{equation}
    
\noindent Image reconstructions abide by two restrictions: the flux in a particular pixel must be non-negative and the flux of the reconstructed image is normalized to unity.

The image reconstruction code SQUEEZE\footnote{https://github.com/fabienbaron/squeeze} is used on the $\lambda$ And data sets \citep{baron10}; this code is an evolved version of the reconstruction program MACIM \citep{ireland06}.  The SQUEEZE initial state is a 50 x 50 pixel array of a 2.777 mas  (the measured $\theta$ for $\lambda$ And) uniform disk containing 4000 flux elements.  After the initial state is set, SQUEEZE uses a multi-thread approach with each thread finding the best image by moving flux from pixel to pixel iteratively via a simulated annealing algorithm.  The \emph{total variation} regularizer is chosen as it is designed to minimize brightness gradients across the surface \citep{rudin92}.  This favors a conservative surface image with a few large starspots as opposed to many smaller ones.  Experiments with other regularizers such as Laplacian regularization and the $\ell_{0}$ sparsity norm were attempted but failed to produce images with reduced $\chi^{2}$ lower than total variation alone. This is consistent with the findings of \citet{renard11} who found total variation outperformed four other regularization methods. 

\begin{figure}
  \plotone{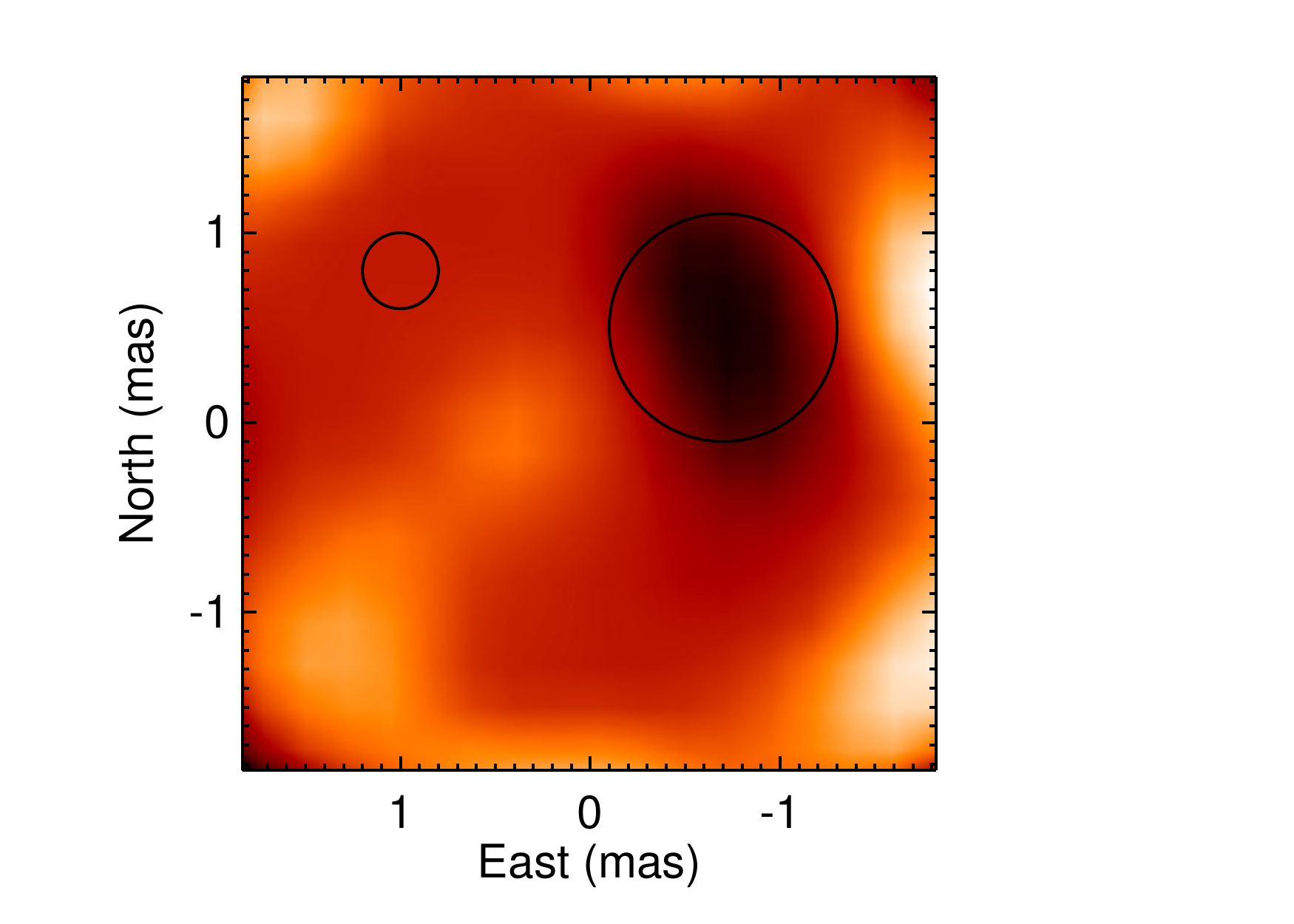}
  \caption{Shown is a closeup of the SQUEEZE reconstruction for the Sep 2$^{th}$, 2011 data near an apparent starspot.  The black circle on the right shows the aperture used to extract starspot properties from reconstructed image.  The black circle on the left shows the aperture over the ``quiet'' photosphere.  The ``quiet'' photosphere is defined as a part of the stellar surface devoid of flux gradients.  The size of the aperture is identical to the minimum achievable angular resolution.}
  \label{fig:detectstr}
\end{figure}

A final image reconstruction is the average of ten images generated by SQUEEZE. The initial conditions in regards to the initial image (uniform disk), choice of regularizer (total variation), and hyper-parameter are identical for each ten images.  The random seed used for the Markov chains to create the image is unique for each image.  Prior to being averaged, the ten images were registered using SPLASH\footnote{https://github.com/fabienbaron/splash} to account for any changes in the image photocenter.  The averaging of ten images attempts to minimize the effect of artifacts potentially caused by the reconstruction process.  Starspot parameters are extracted by fitting a circular aperture over identified starspots.  This process is illustrated in Fig~\ref{fig:detectstr}.  The aperture size provides the covering factor and the location of the aperture center provides the starspot latitude and longitude.  As the starspot edge is difficult to quantify and the starspot may be irregular in shape, $\phi$ is considered a lower bound.  The intensity ratio is calculated by dividing the flux at the aperture center with a flux measurement of the ``quiet'' photosphere.  The quiet photosphere is identified as a part of the stellar surface devoid of flux gradients.  The circular aperture is fit to the reconstructed starspots by eye.

In addition to the creation of a final averaged reconstructed image, an image representing the standard deviation of the ten iterations is created.  The detection strength, $\sigma_{ds}$, of the starspot is computed using this mean standard deviation.  A circular aperture is placed on the quiet photosphere with a size equal to the minimum angular resolution.  The detection strength is the mean flux within this aperture subtracted from the mean flux within the starspot aperture and then divided by the standard deviation.

\subsubsection{Image Artifacts}
\label{subsubsec:artifacts}

A large problem facing image reconstruction is the believability of surface features.  Certain results from imaging can be easily dismissed as artifacts such as hexagonal stellar surfaces or a repeating symmetric patterns of bright spots that is inconsistent with a rotating surface.  Certain results can be accepted as believable such as dark starspots common between the reconstructed and modeled images due to the independence between the two imaging methods.  However, since single bright starspots were not modeled, this test is unavailable.  Therefore a method must be devised to determine if any imaged bright starspots are real (e.g. flares, plages) or a consequence of the data sampling and/or image reconstruction process.  

The method used in this analysis is to create synthetic reconstructed images based on the model images.  This is done by extracting visibilities, triple amplitudes, and closure phases from each model image with identical data sampling and SNR as the measured data using the OIFITS-SIM tools\footnote{https://github.com/bkloppenborg/oifits-sim}.  The same reconstruction procedure described in $\S$~\ref{subsec:squeeze} is used to create synthetic reconstructed images from these data.  Artifacts due to miscalibrated observables will be features seen in the reconstructed image, but are absent in both the simulated and the model images.  Artifacts due to the [u,v] coverage and the reconstruction process will be features seen in both the reconstructed and simulated images, but not in the true image.  

\section{DISCUSSION}
\label{sec:disc}

\subsection{$\lambda$ Andromedae Stellar Properties}
\label{subsec:theta_alpha}

\begin{figure}
  \plotone{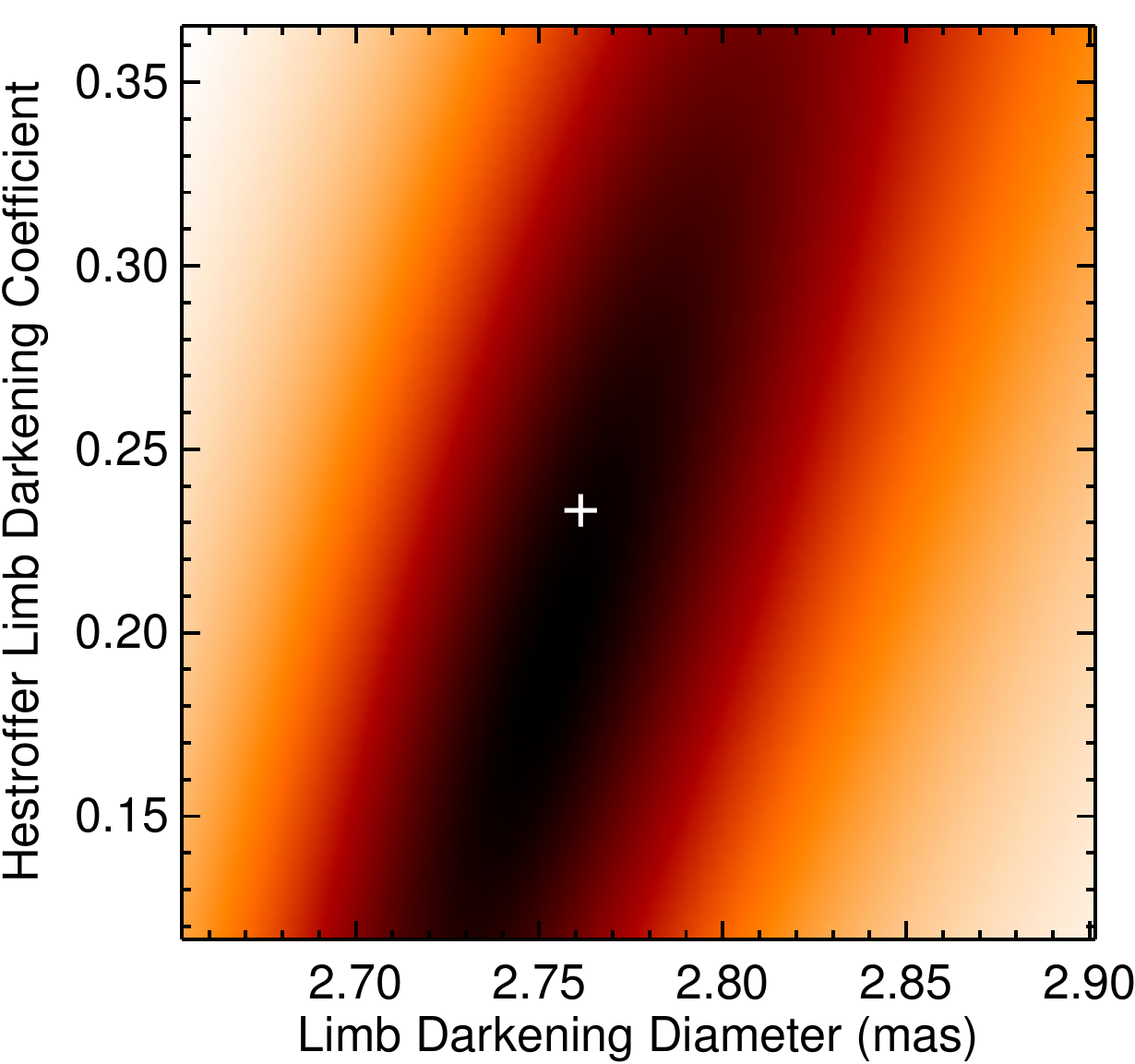}
  \caption{The best fit $\theta$ and $\alpha$ are determined through a grid search with a parameter step size of $1\times 10^{-3}$.  The reduced $\chi^{2}$ space for $\theta$ and $\alpha$ is shown with the white cross marking the position of best fit.  A slight degeneracy exists between these two parameters as seen by the elliptically-shaped gradients.}
  \label{fig:thealp_error}
\end{figure}

The angular diameter and limb-darkening coefficient are determined via the modeling described in $\S$~\ref{subsec:spot_model}.  The initial value of $\theta$ for the AMOEBA code is set to 2.75 mas as determined from the $\lambda$ And (\emph{V}-\emph{K$_{s}$}) color \citep{vanbelle99}.  An initial $\alpha$ is found by matching a power-law fit to a four parameter fit from \citet{claret11} given the coefficients for a star with \emph{$T_{\mathrm{eff}}$} = 4750 K and log(g) = 3.0 dex (cgs).  This yielded a result of $\alpha$ = 0.22 consistent with results from other power law fits to interferometric data of late type giants \citep{wittkowski02,wittkowski06}.  The search scales were roughly 10$\%$ of the initial values.  The final results are $\theta$ = 2.759 $\pm$ 0.050 mas and $\alpha$ = 0.229 $\pm$ 0.111.  The uncertainty is measured from the lowest $\chi^{2}$ + 1 ellipse in the reduced $\chi^{2}$ space.  The error bars correspond to the furtherest points of the ellipse along each axis, respectively.  Fig~\ref{fig:thealp_error} shows the reduced $\chi^{2}$ space for $\theta$ and $\alpha$ with the white cross marking the best solution.  A slight degeneracy exists between these two parameters as seen by the elliptically-shaped gradients.  At a \emph{Hipparcos} trigonometric distance of 37.87 $\pm$ 0.21 mas \citep{vanleeuwen07}, this angular diameter corresponds to a linear radius of 7.831$^{+0.067}_{-0.065}$ R$_{\sun}$. 

\subsection{$\lambda$ Andromedae Starspot Properties: 2010 Data Set}
\label{subsec:2010}

\begin{figure}
  \plotone{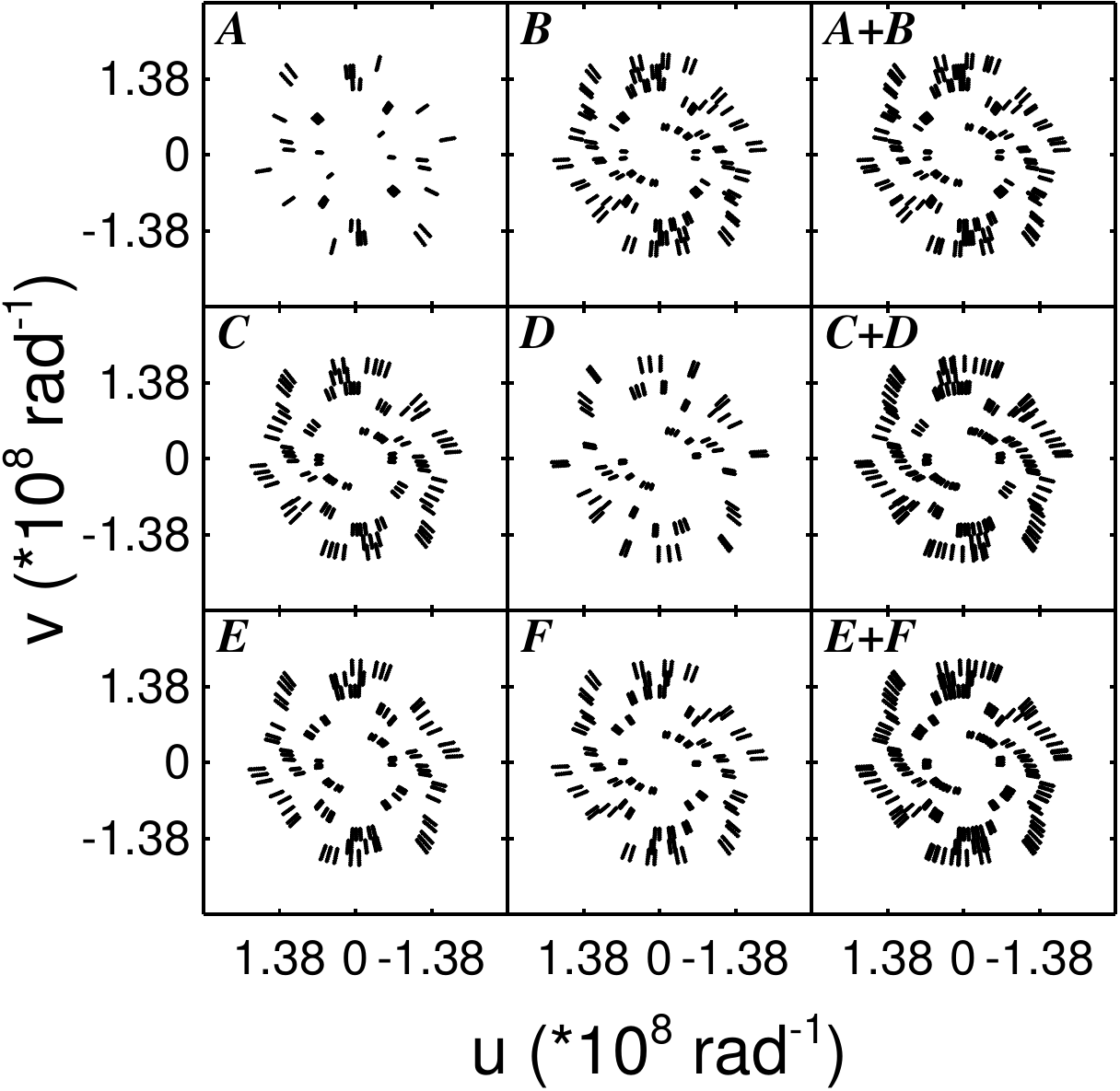}
  \caption{The [u,v] coverage obtained for the first three epochs in 2010.  A - Aug 2$^{nd}$; B - Aug 3$^{rd}$; C - Aug 10$^{th}$; D - Aug 11$^{th}$; E - Aug 18$^{th}$; F - Aug 19$^{th}$}
  \label{fig:uvplottwo}
\end{figure}

\begin{figure}
  \plotone{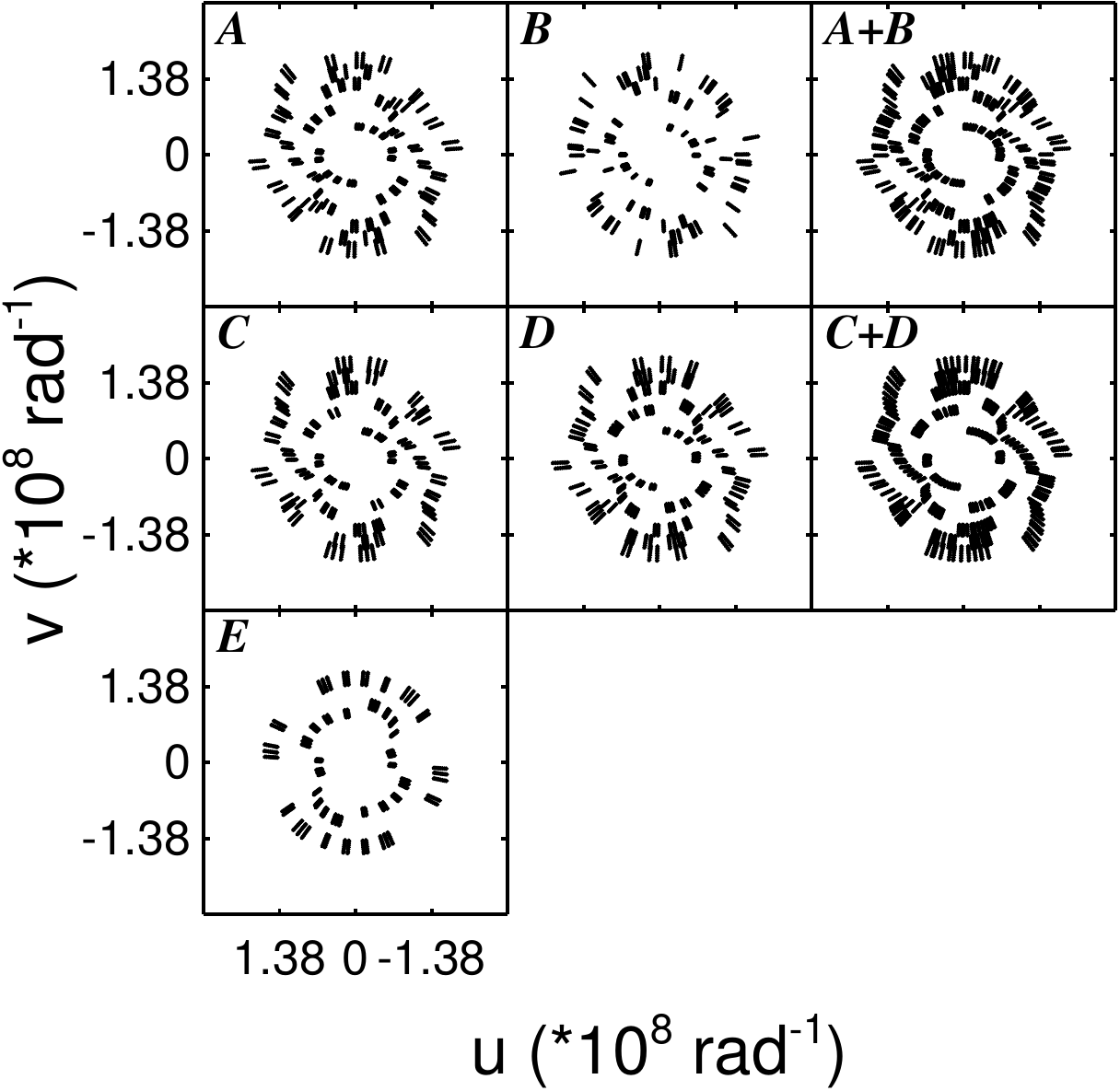}
  \caption{The [u,v] coverage obtained for the second three epochs in 2010.  A - Aug 24$^{th}$; B - Aug 25$^{th}$; C - Sep 2$^{nd}$; D - Aug 25$^{th}$; E - Sep 10$^{th}$}
  \label{fig:uvplotthree}
\end{figure}

Between Aug 2$^{nd}$ and Sep 11$^{th}$, 2010, 6 epochs of data were obtained via the strategy described in $\S$~\ref{subsec:int_obs}.  The number of [u,v] points obtained for each of the two night combined 5 epochs ranged from 624 to 1128 with the densest converge obtained by the combination of Sep 2$^{nd}$ and 3$^{rd}$ (see Table~\ref{tab:obs_log}).  The number of [u,v] points for the single night epoch on Sep 10$^{th}$ is 336.  Fig~\ref{fig:uvplottwo} and Fig~\ref{fig:uvplotthree} show the distribution of [u,v] coverage obtained for each pair of observations.  The six epochs are spaced with a cadence between six to nine days corresponding to 10.9$\%$ to 16.4$\%$ of the measured rotation period; significant apparent starspot motion is expected between epochs.  The complete observing run spans 71$\%$ of one complete $\lambda$ And rotation period.  

\begin{figure}
  \plotone{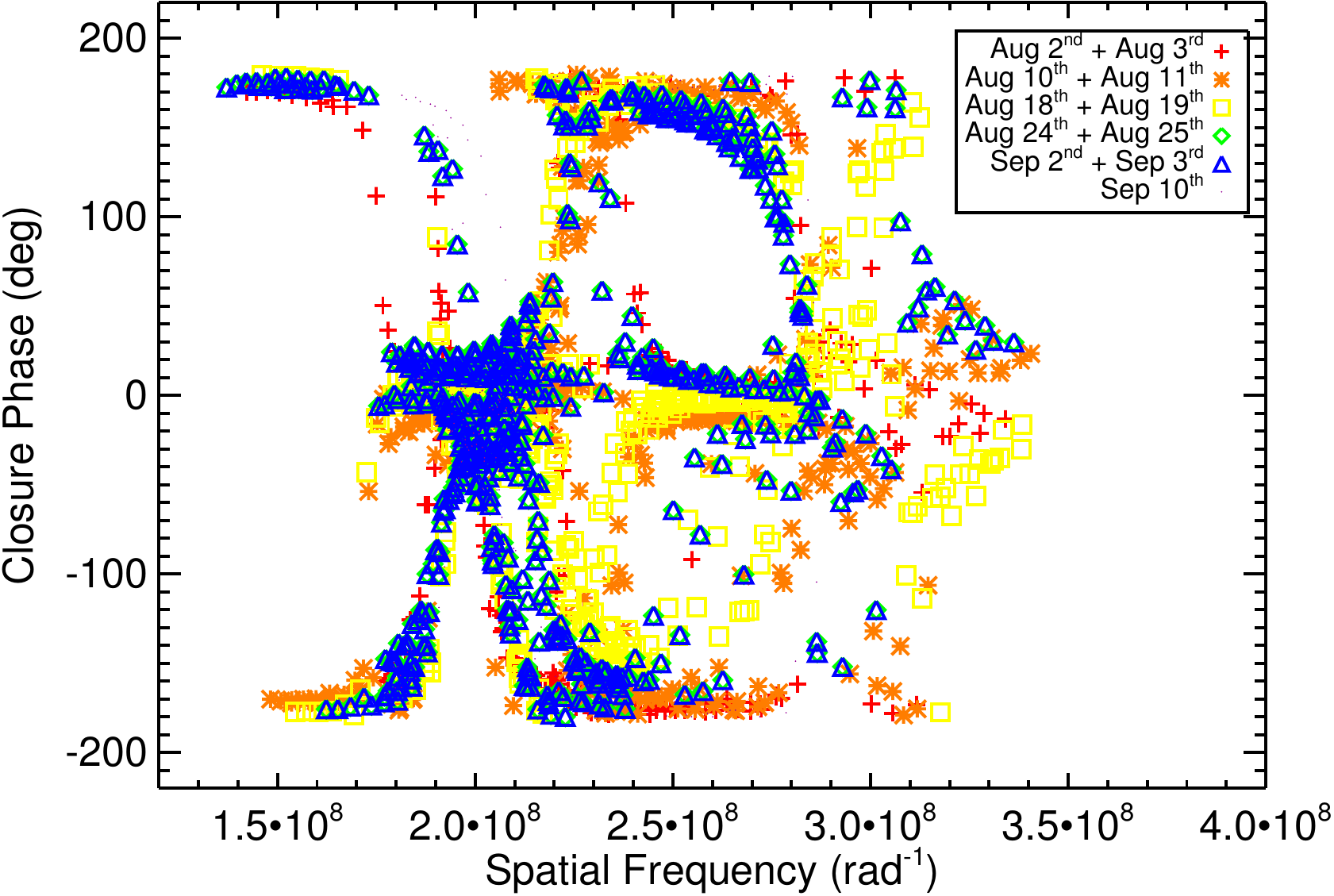}
  \caption{The observed closure phases for the 2010 data set.  The errors bars have been excluded for clarity. The distinct non-zero closure phase signature points to surface asymmetries.  The differences in the closure phase between nights indicates an evolving asymmetric surface pattern from night to night.}
  \label{fig:tencp}
\end{figure}

Fig~\ref{fig:tencp} shows a distinct non-zero closure phase signature across most sampled spatial scales pointing to the existence of surface asymmetries.  This signature is present in all six epochs.  The measured closure phase distribution differs between epochs supporting the hypothesis that the apparent starspot configuration evolves as the star rotates.  In addition, an unspotted model image does not fit the interferometric data for each epoch in 2010 with the reduced $\chi^{2}$ ranging between 15.4 and 44.0.

\begin{figure}
  \plotone{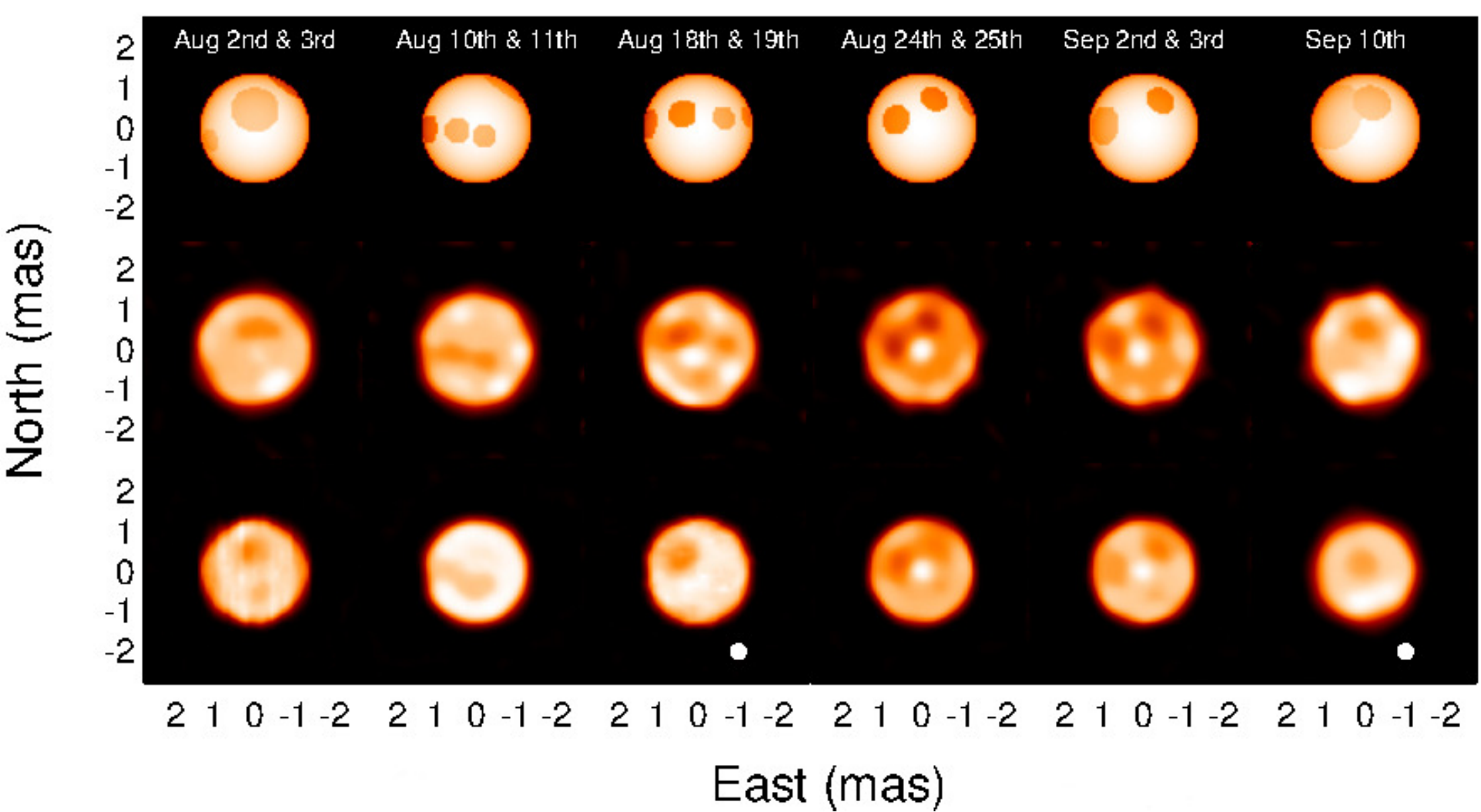}
  \caption{Stellar surface images for the 2010 data set. The top row contains the model images, the middle row contains the reconstructed images, and the bottom row contains the simulated images.  The white dot in the lower right hand corner represents the 0.4 mas resolution limit for the CHARA array.}
  \label{fig:tendataplot}
\end{figure}

The best-fit models for each epoch contain between two to four cool starspots.  Fig~\ref{fig:tendataplot} contain the model, reconstructed and simulated images for each epoch.  The model reduced $\chi^{2}$ range between 2.97 to 6.02 with the best fit occurring for Aug 10$^{th}$ and 11$^{th}$.  As an ensemble, the covering factor for individual starspots, $\phi$, ranges from 4.0 to 21.8$\%$ with a median value of 7.6$\%$.  The covering factor errors range 2.6 to 5.5$\%$ with a median error of 4.1$\%$.  The temperature ratio, \emph{T$_{R}$}, ranges from 0.756 to 0.926 with a median value of 0.853.  The temperature ratio errors range from 0.007 to 0.042 with a median error of 0.015.  The errors in both latitude and longitude are nearly identical and range from 0.75 to 7.8$\degree$ with a median error of 1.9$\degree$.

The reduced $\chi^{2}$ for each reconstructed image is $\leq$ 1.01.  Good qualitative agreement exists between the model and reconstructed images (see Fig~\ref{fig:tendataplot}).  Table~\ref{tab:tendata} contains the measured starspot parameters from both the model and reconstructed images.  This table also contains the difference between the parameters for the same starspot in the model and reconstructed images.

The Epoch 6 model image contains two cool starspots.  Although both starspots are listed in Table~\ref{tab:tendata}, the starspot located near the northeastern limb (\emph{b}: 17.5$\degree$, \emph{l}: -40.9$\degree$) is excluded when discussing ensemble starspot properties.  The rationale for this decision is: the starspot is nearly twice the size ($\phi$ = 44$\%$) of the next largest modeled starspot, it is the warmest starspot (T$_{R}$ = 0.925), and is not confidently seen in the reconstructed image.  This ``starspot'' could be a widely spread patch of starspots with the covering factor of each individual starspot falling below the resolution limit .  The model attempts to reconcile the interferometric signature these starspots produce through the addition of this larger, warmer starspot.

The reconstruction process does not identify starspots on the stellar limb nearly as well as the model.  For epochs with visual agreement between model and reconstructed starspots, $\phi$ and T$_{R}$ estimates are within one error bar in nearly every case.  The agreement suffers for position estimates with starspots in reconstructions consistently further east and/or south than their model image counterparts.

The simulated images and the observing cadence are used to help identify potential artifacts in the reconstructed images.  In Epochs 2 through 6, a number of warm starspots are observed in the reconstructed image evenly spaced around the stellar limb.  These are rejected as artifacts due to their symmetry and constant position despite the rotating surface.  The origin of these artifacts may be due to the [u,v] sampling since the pattern of the warm starspots is similar to the pattern of tightly clustered points in the [u,v] plane (see Fig~\ref{fig:uvplotthree}).  The southern cool starspot in Epoch 2 is rejected as the covering factor is below the resolution limit.  The central warm starspot in the reconstructed images of Epochs 3 and 4 are rejected as artifacts; they are reproduced in the simulated images and are contrary to a rotating surface.  The brighter southern pole in Epoch 6 is reproduced in the simulated image and thus rejected.  The warm southwestern starspot Epoch 1, however, cannot be rejected as false as it is not reproduced.

\subsubsection{Comparison with the 2010 light curve}
\label{subsubsec:2010_photcomb}

Determining if the starspot imaging provides a consistent picture with the measured photometric time series will provide further confidence in the interferometric results.  Season 2010 spans 121.8 days and has peak-to-trough photometric $\Delta$V amplitude of 0.099 mag.  The time series begins 10 days after the interferometric observations providing no simultaneous coverage.  The seasons spans $\sim$2.2 rotation periods, and varies from one rotation to the next indicating starspot evolution between rotations.  Also the light curve seems bimodal suggesting starspots on two active longitudes ($\phi$ = 0.0 and 0.4).  Active longitudes, or longitudes of preferential starspot formation, have been associated with magnetically active stars \citep{berdyugina05}.  This could be an example of active longitude migration with respect to the star's rotational frame of reference \citep{jetsu93,lanza98,berdyugina98}.

\begin{figure}
  \plotone{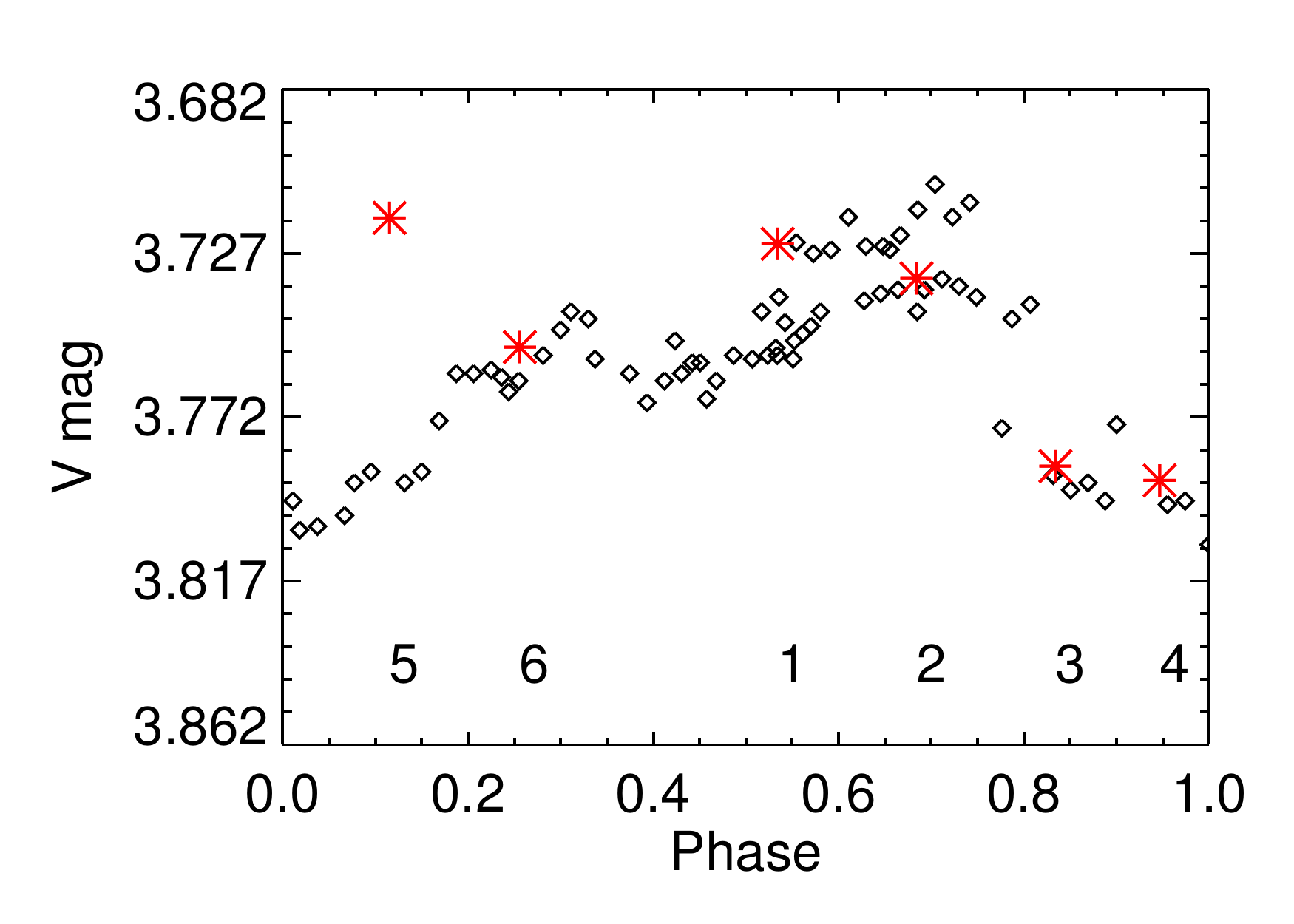}
  \caption{The gray diamonds correspond to the \emph{V}-Band time series phased to a period of 53.3 $\pm$ 1.9 days.  The red asterisks represent the photometry taken from the best-fit models for the 6 epochs. The numbers along the bottom axis indicate the epoch.}
  \label{fig:lcten}
\end{figure}

A modeled light curve can be constructed from the best-fit models for each epoch.  The change in intensity between an unspotted star and the modeled surface is measured, converted into a $\Delta$ magnitude and then scaled for comparison with the observed time series.  The scaling is done through an additive constant that shifts the modeled time series to the approximate values of the photometric time series.  A multiplicative constant is used to expand the variability amplitude to be comparable to the photometric time series.  A modeled light curve is computed and shown in Fig~\ref{fig:lcten} overlaid on top of the photometric time series folded to the 53.35 day period.  There is good visible agreement in all epochs except of Epoch 5, which is brighter than expected.  It is difficult to explain the discrepancy based on [u,v] coverage since this epoch had the densest coverage.  Data quality does not seem to be a viable explanation as the errors are not significantly larger than other epochs and the model reduced $\chi^{2}$ is one of the lowest all six epochs.

\subsubsection{Tracing Rotation in the 2010 Data Set}
\label{subsubsec:trace2010}

Multi-epoch starspot imaging has the potential to trace the stellar rotation via apparent starspot motion.  If this motion can be observed, a rotation period can be measured and compared to the period derived from photometry.  In addition, the stellar rotation axis can be fully described in both inclination and position angle.  Neither Doppler imaging or light curve inversion has the capacity to determine the latter two quantities and, in fact, the inclination angle must be assumed in both cases.

The observing baseline for the 2010 data set spans 71$\%$ of the photometrically determined rotation period.  The cadence will carry starspots $\sim$47$\degree$ across the stellar surface between epochs assuming a negligible amount of differential rotation.

\begin{figure}
  \plotone{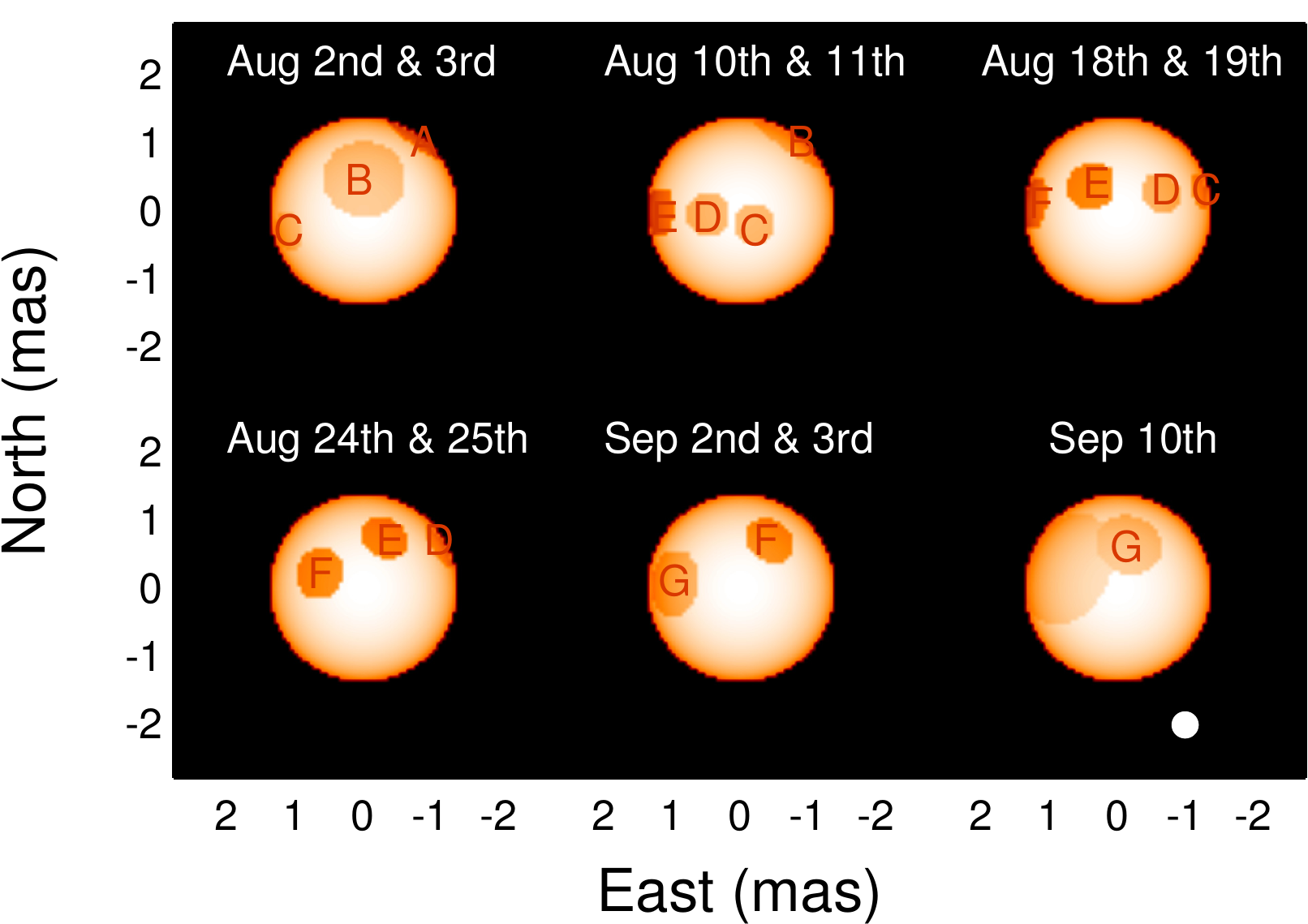}
  \caption{The best fit models for each epoch in 2010.  In each model, the starspot(s) are labeled (\emph{A} through \emph{G}) to indicate the same starspot as seen in each rotational phase.}
  \label{fig:tenspotrot}
\end{figure}

Individual starspots identified in different epochs are labeled \emph{A} through \emph{G} in Fig~\ref{fig:tenspotrot}.  Identification of the same starspot in different epochs is judged via visual inspection.  Four starspots (\emph{C}, \emph{D}, \emph{E}, \emph{F}) are believed to be seen in three epochs and therefore provide the most useful constraints on the rotation and inclination angle.  Changes in covering factor and temperature ratio are described in Table~\ref{tab:rotation10}, along with the computed rotation period based on the measured angle between starspot positions from one epoch to the next.  The change in $\phi$ and \emph{T$_{R}$} should be small since these starspot characteristics are not expected to evolve during one stellar rotation.  The median $\Delta\phi$ and $\Delta$\emph{T$_{R}$} are 0.5$\%$ and 0.014, respectively.  As these are below the median $\phi$ error of 4.6$\%$ and median \emph{T$_{R}$} of 0.016, it is safe to claim that these starspots are not significantly evolving.  The range in measured rotation periods is 44.3 to 78.7 days with an average rotation period of 60 $\pm$ 13 days; the error is the standard deviation of the individual measured rotation periods.  The starspot determined period is consistent with the photometric rotation period of 54.02 $\pm$ 0.88 days due to the large error bar.   The large spread in starspot determined rotation periods is likely a consequence of four starspots observed only three times over an incomplete rotation. 

\begin{figure}
  \plotone{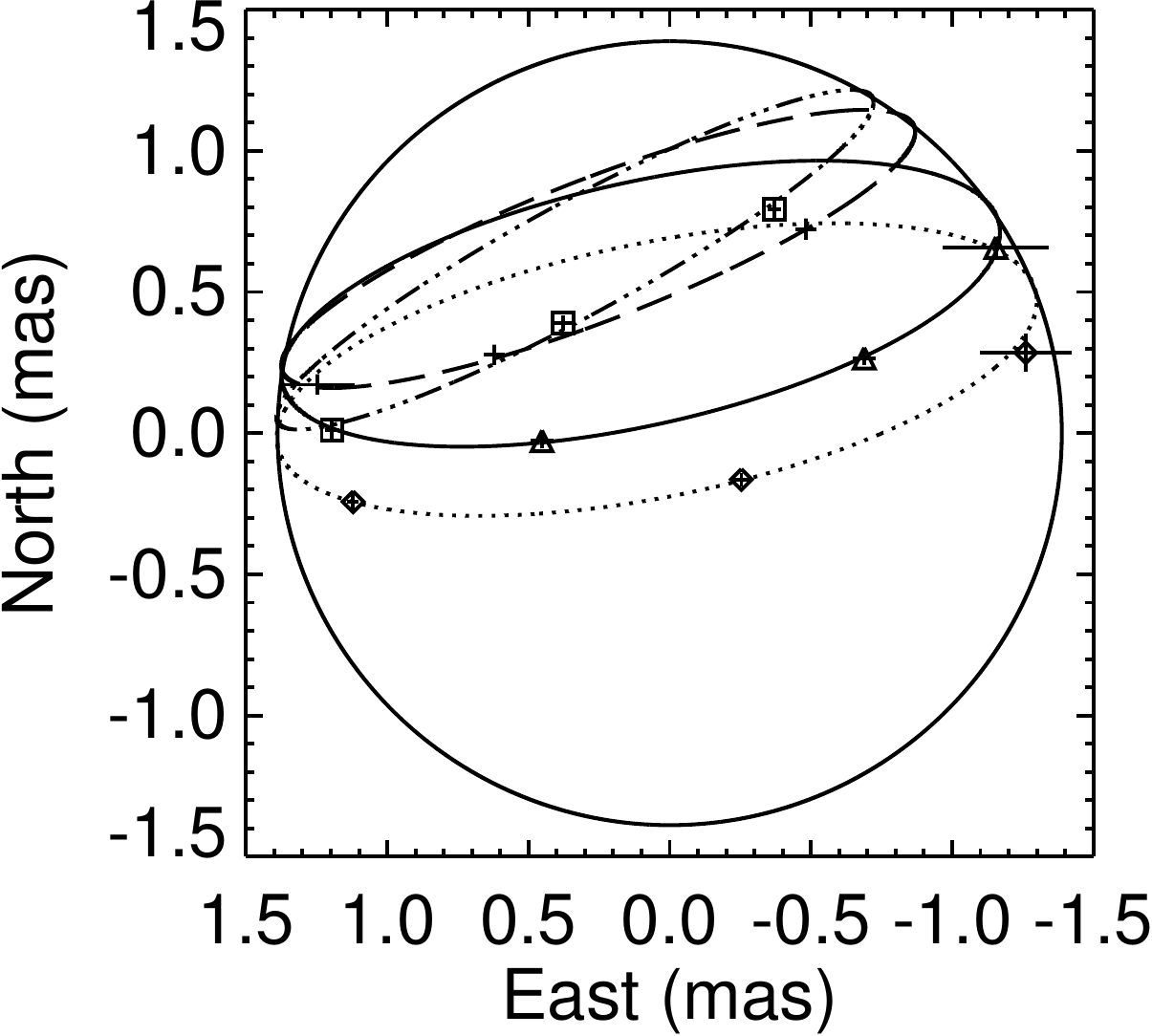}
  \caption{Ellipse fits to starspot positions in the 2010 data sets.  The dotted line corresponds to spot \emph{C}.  The solid line corresponds to spot \emph{D}.  The dot dashed line corresponds to spot \emph{E}.  The dashed line corresponds to spot \emph{F}.  The average computed position angle, $\Phi$, and inclination angle, \emph{i}, from these fits are 19 $\pm$ 8.1$\degree$ and 75 $\pm$ 5.0$\degree$, respectively.  The solid line circle corresponds to the circumference of $\lambda$ And.}
  \label{fig:tenposang}
\end{figure}

Fig~\ref{fig:tenposang} shows each of the starspots plotted by longitude vs. latitude overlaid by ellipse fits.  Estimations of the inclination and position angle of the rotation angle can be made by measuring these elliptical paths.  A starspot being carried across the stellar surface via rotation will appear to travel along an elliptical path when viewed in two dimensions.  The position angle, $\Psi$, is simply the tilt of this ellipse counterclockwise from north (up).  The inclination angle, \emph{i}, is the inverse sine of the ellipse eccentricity.  If the star is viewed face-on, then the starspot will appear to traverse a circular path.  Conversely if the star is viewed edge-on, the starspot will appear to traverse a line.  An ellipse is fit via visual inspection to starspots \emph{C} through \emph{F} as these spots have been identified in at least 3 epochs.  The average $\Psi$ and \emph{i} are 19 $\pm$ 8.1$\degree$ and 75 $\pm$ 5.0$\degree$, respectively.  The errors are the standard deviations of measured values.  This inclination angle is higher than the 60$\degree$ assumed by F08, but is consistent within the uncertainties in the previous inclination estimate 60$^{+30}_{-15}$$\degree$ by D95.  The rotation axis is measured to be coming out of the plane of the sky in the northern hemisphere.

\subsection{$\lambda$ Andromedae Starspot Properties: 2011 Data Set}
\label{subsec:2011}

\begin{figure}
  \plotone{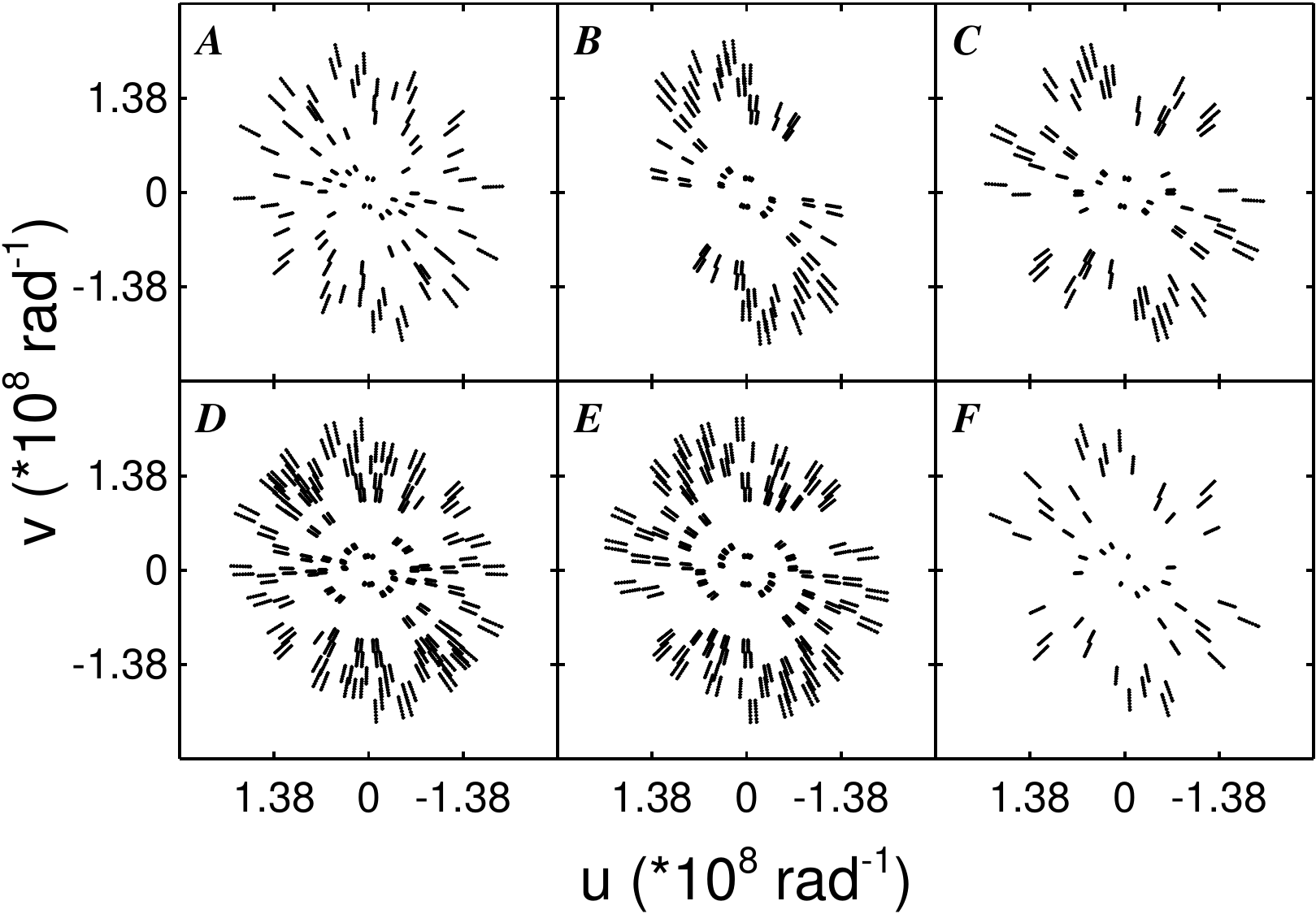}
  \caption{The [u,v] coverage obtained for the 2011 observing run.  A - Sep 2$^{nd}$; B - Sep 6$^{th}$; C - Sep 10$^{th}$; D - Sep 14$^{th}$; E - Sep 19$^{th}$; F - Sep 24$^{th}$}
  \label{fig:uvplotfour}
\end{figure}

Between Sep 2$^{nd}$ and Sep 24$^{th}$, 2011, six epochs of data were obtained for $\lambda$ And using all six telescopes simultaneously.  In order to maximize the [u,v] coverage, observations were taken for as long as delay lines were available.  Only one night of data was acquired per epoch as opposed to the combined days in 2010.  Table~\ref{tab:obs_log} contains the number of [u,v] obtains points per observation.  The number of [u,v] points achieved ranges from 200 to 864 with the densest coverage obtained on Sep 14$^{th}$ (see Fig~\ref{fig:uvplotfour}).  The observing cadence was four or five days corresponding to 7.3$\%$ and 9.2$\%$ of the rotation period; all six observations span 40.4$\%$ of one rotation period.   

\begin{figure}
  \plotone{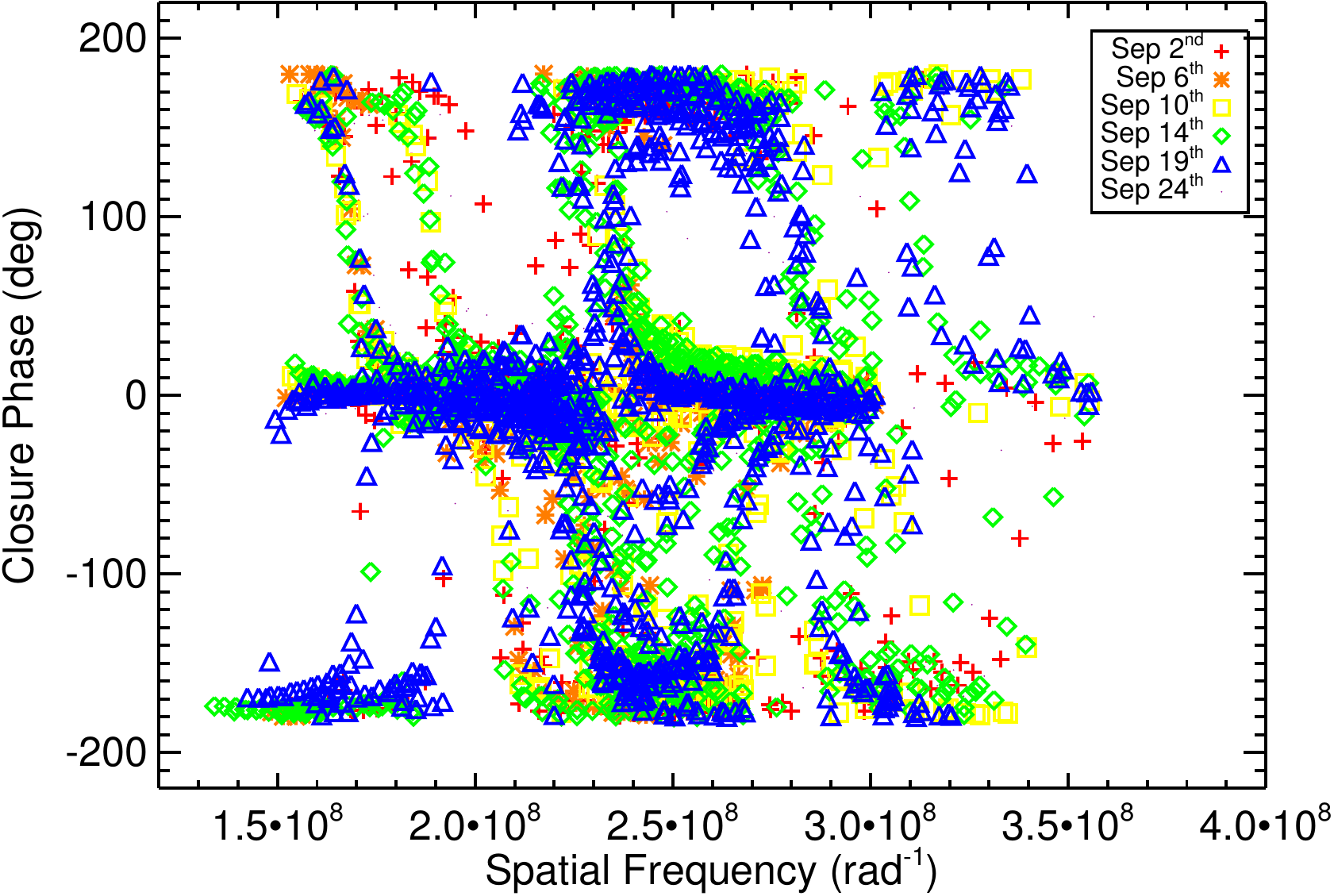}
  \caption{The observed closure phases for the 2011 data set.  The error bars have been removed for clarity.  The distinct non-zero closure phase signature points to surface asymmetries.  The differences the closure phase between nights indicates an evolving asymmetric surface pattern from night to night.}
  \label{fig:elevencp}
\end{figure}

Fig~\ref{fig:elevencp} shows a distinct non-zero closure phase signature across most sampled spatial scales in all six epochs pointing to an evolving configuration of surface asymmetries.  In addition, an unspotted model image does not fit the interferometric data for each epoch in 2011 with the reduced $\chi^{2}$ ranging between 9.3 to 16.7.

\begin{figure}
  \plotone{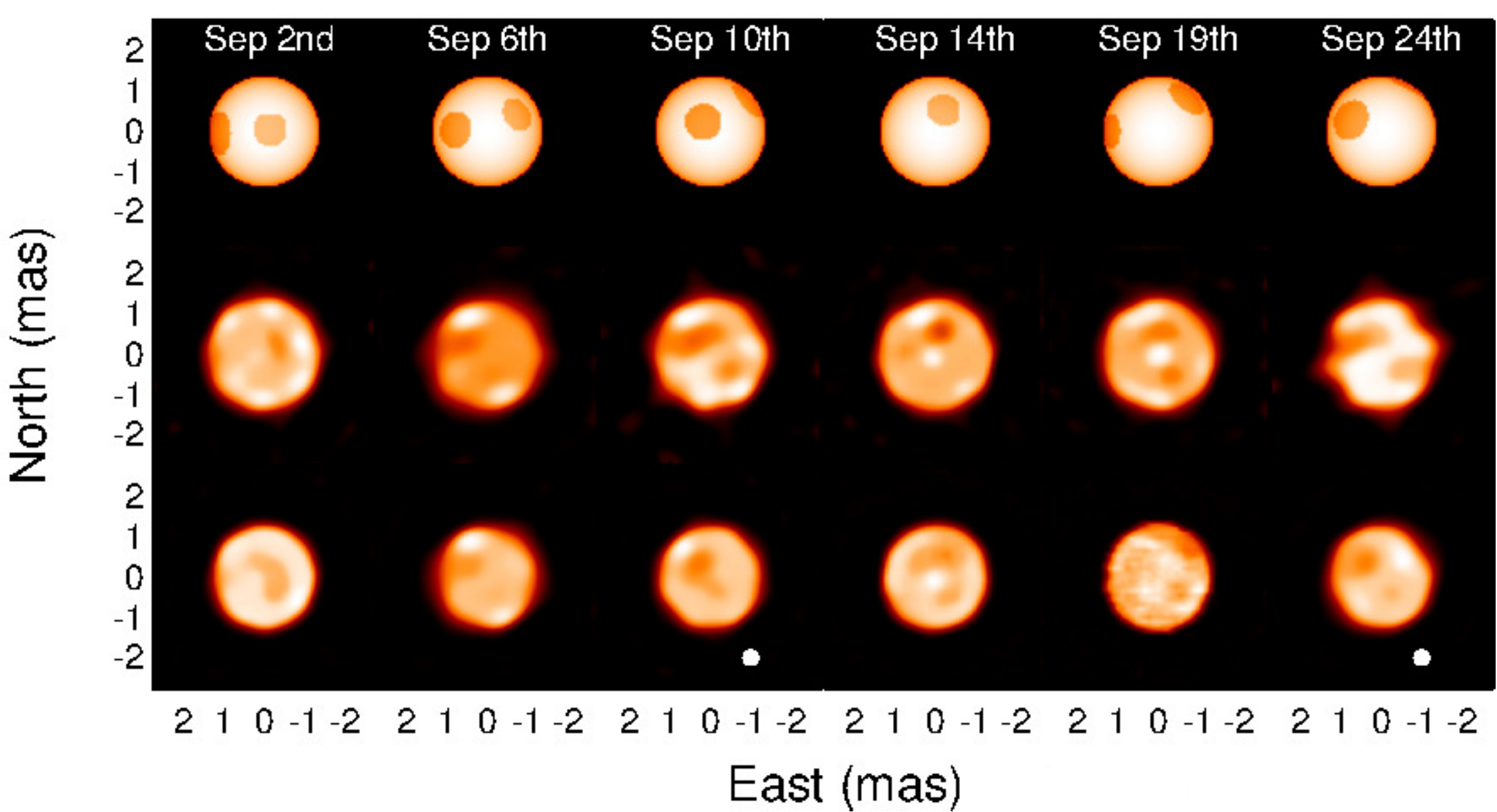}
  \caption{Stellar surface images for the 2011 data set. The top row contains the model images, the middle row contains the reconstructed images, and the bottom row contains the simulated images.  The white dot in the lower right hand corner represents the 0.4 mas resolution limit for the CHARA array.}
  \label{fig:elevendataplot}
\end{figure}

The best-fit models for each epoch contain one or two cool starspots.  The model reduced $\chi^{2}$ range between 3.14 to 6.57 for these epochs with the best fit occurring for Sep 2$^{nd}$.  The starspot properties are listed in Table~\ref{tab:elevendata}.  Fig~\ref{fig:elevendataplot} contain the model, reconstructed and simulated images for each epoch. As an ensemble, the covering factor, $\phi$, ranges from 10$\%$ to 17$\%$ with a median value of 12$\%$.  The errors in these values range from 2.5$\%$ to 6.4$\%$ with a median error of 5.3$\%$.  The temperature ratio, \emph{T$_{R}$}, ranges from 0.797 to 0.867 with a median value of 0.842.  The temperature ratio errors range from 0.007 to 0.016 with a median error of 0.014.  The errors in both latitude and longitude are nearly identical and range from 0.63 to 5.7$\degree$ with a median error of 1.4$\degree$.  

The reduced $\chi^{2}$ for each of the reconstructed images is at or below 1.01.  Good qualitative agreement exists between the model and reconstructed images, however some exceptions do exist (see Fig~\ref{fig:elevendataplot}).  It is unclear why the western starspot seen in model images is not recovered in the reconstructed images for Sep 6$^{th}$ and 10$^{th}$.  Nor is it clear why there is poor agreement between the Sep 19$^{th}$ model and reconstructed images.  In general, the agreement between starspot properties is not as good as it is in the 2010 data set.  The reconstructed covering factor is always smaller than the modeled covering factor in each epoch, however, this is not surprising as this parameter should be considered a lower bound.  In an opposite trend than 2010, the reconstructed \emph{b} is further north where it deviates from the model image.  No trend exists for when the agreement in \emph{l} exceeds the uncertainties.

The reconstructed images in Epochs 1, 3, and 5 contain a ring of warm starspots around the stellar limb.  These starspots are rejected as artifacts caused by [u,v] sampling due to their symmetry and constant location between epochs, which is contrary to starspots on a rotating surface.  The following starspots are rejected as artifacts by the presence of similar features in the simulated image.  In Epochs 2 and 4, the warm starspots in the northeast and southwest are rejected as artifacts.  The warm northeast starspot in Epoch 3, the warm central starspot in Epoch 4, and the cool southern starspot in Epoch 5 are all rejected as artifacts.  The non-circular stellar disk in Epoch 6 is most certainly an artifact due to the limited number of [u,v] points (200).  In addition, the shape of this disk resembles the configuration of [u,v] points (see Fig~\ref{fig:uvplotfour}).

\subsubsection{Comparison with the 2011 light curve}
\label{subsubsec:2011_photcomb}

Again the measured photometric time series is compared to the imaging results to provide further confidence in the interferometric results. Season 2011 spans 134.9 days and has a peak-to-trough $\Delta$\emph{V} magnitude of 0.057 mag.  This season spans $\sim$2.5 rotation periods and overlaps the first two epochs of interferometric observations.  The folded light curve is double-peaked; this is potentially due to two active longitudes separated by $\sim$180$\degree$.

\begin{figure}
  \plotone{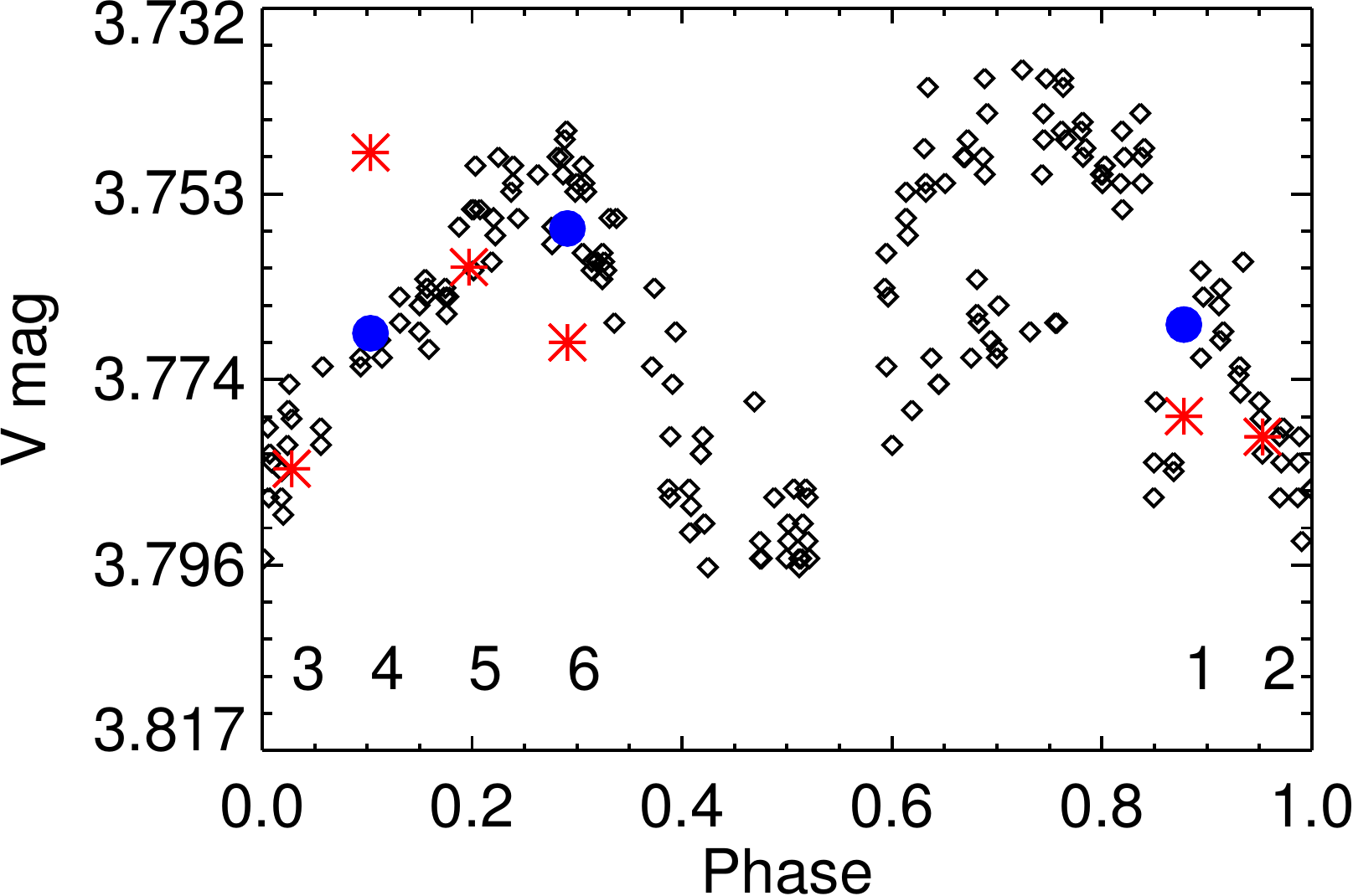}
  \caption{The gray diamonds correspond to the \emph{V}-Band time series phased to a period of 53.3 $\pm$ 1.9 days.  The red asterisks represent the photometry taken from the best-fit models for the 6 epochs.  The filled blue circles represent model photometry where the values of $\phi$ and/or T$_{R}$ have been altered (see text).}
  \label{fig:lceleven}
\end{figure}

A modeled light curve is computed and shown in Fig~\ref{fig:lceleven} overlaid on top of the photometric time series folded to the 53.30 day period.  The modeled photometry in each epoch is given by the red asterisks and is agreement with the observed photometry in Epochs 2, 3, and 5.  The agreement for Epoch 1 is not as convincing, but not completely inconsistent.  However, the modeled photometry for Epochs 4 and 6 visibly disagrees with the observed photometry.  The filled blue circles represent the modeled photometry, if the parameters for a single starspot are changed.  The altered starspots are \emph{B} in Epochs 1 and 4 and \emph{C} in Epoch 6.  The starspot covering factor in both Epochs 1 and 4 is lowered to 11.8$\%$, which is identical to that of Epoch 3 where the starspot is most centrally located.  Additionally the Epoch 4 starspot T$_{R}$ is lowered to that of starspot \emph{B} in Epoch 3 (0.797).  The starspot covering factor in Epoch 6 is lowered to 10.4$\%$, which is the same as the $\phi$ for the same starspot in Epoch 5. 

\subsubsection{Tracing Rotation in the 2011 Data Set}
\label{subsubsec:trace2011}

\begin{figure}
  \plotone{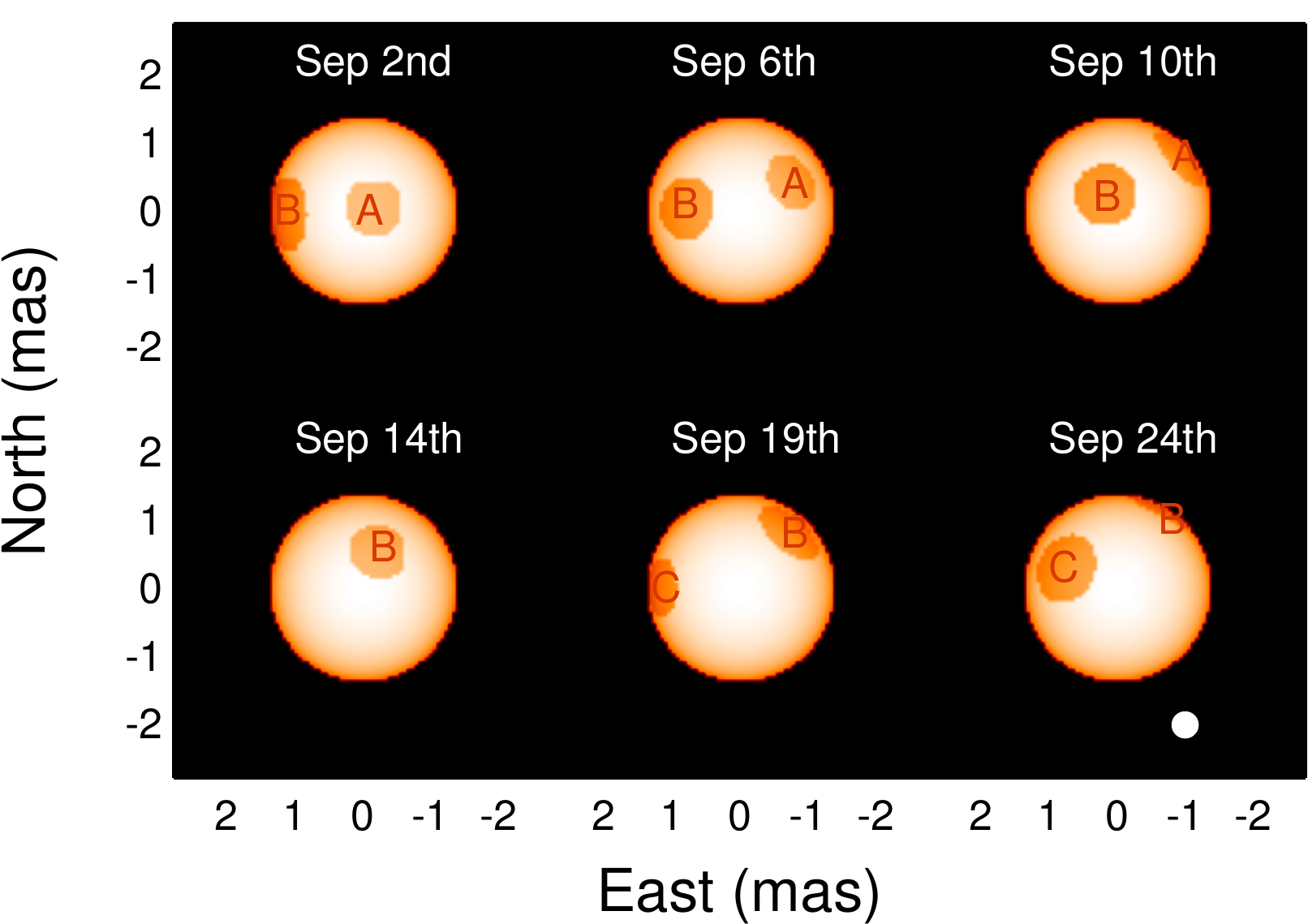}
  \caption{The best fit models for each night in 2011.  In each model, the starspot(s) are labeled (\emph{A}, \emph{B}, and \emph{C}) to indicate the same starspot as seen in each rotational phase.}
  \label{fig:elevenspotrot}
\end{figure}

The observing baseline for the 2011 data set spans $\sim$41$\%$ of the photometrically determined rotation period.  The cadence will carry starspots $\sim$7$\%$ across the stellar surface between epochs assuming a negligible amount of differential rotation.  Fig~\ref{fig:elevenspotrot} shows a compelling pattern of stellar rotation by three starspots labeled \emph{A} through \emph{C}.  Starspot \emph{B} is seen in all 6 epochs and provides the best estimates of both the stellar rotation and rotation axis.  The observing strategy behind the 2011 data set was designed to provide an increased number of measurements for any individual transiting starspot(s).  This is done to shrink the uncertainties in the estimates of the rotation axis computed from the 2010 data set.  The uncertainties arose from a sparse number of measures per transiting starspot. 

Table~\ref{tab:rotation11} contains the changes in $\phi$ and T$_{R}$ for starspots \emph{A}, \emph{B}, and \emph{C} and the computed rotation period based on the measured angle between starspot positions from one epoch to the next.  The largest change in $\phi$ and \emph{T$_{R}$} for any starspot over the observed rotation is 4.6$\%$ and 0.063, respectively.  As these are comparable to the median $\phi$ error of 5.3$\%$ and median \emph{T$_{R}$} of 0.014, the starspots do not appear to be significantly evolving.  The range in the measured rotation period based on apparent starspot motion is 43.4 to 64.6 days with an average rotation period of 54.0 $\pm$ 7.6 days, which is nearly identical to the photometric rotation period of 54.02 $\pm$ 0.88 days.  The error is the standard deviation of the individual rotation periods. 

\begin{figure}
  \plotone{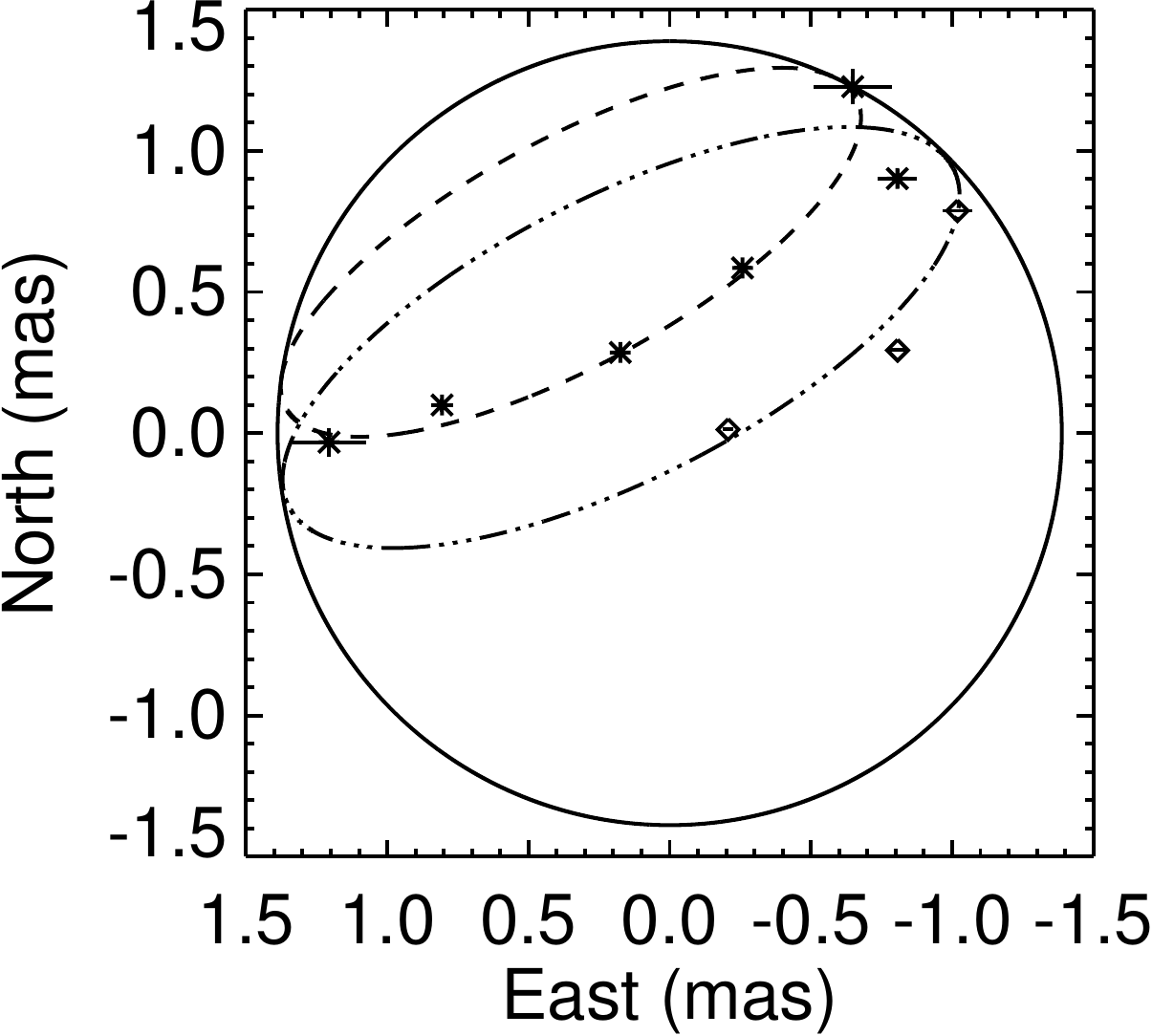}
  \caption{Ellipse fits to starspot positions in the 2011 data sets.  The dashed line corresponds to the fit to spot \emph{A}.  The dash dot dot line corresponds to spot \emph{B}.  The average computed position angle, $\Phi$, and inclination angle, \emph{i}, from these fits are 28 $\pm$ 1.2$\degree$ and 69 $\pm$ 1.4$\degree$, respectively.  The solid line circle corresponds to the circumference of $\lambda$ And.}
  \label{fig:elevenposang}
\end{figure}

Fig~\ref{fig:elevenposang} shows each of the starspots plotted by latitude vs. longitude overlaid by ellipse fits.  For starspots \emph{A} and \emph{B}, an ellipse is fit via visual inspection.  The \emph{C} starspot is excluded from this analysis due to having only two data points.  The average $\Psi$ and \emph{i} are 28 $\pm$ 1.2$\degree$ and 69 $\pm$ 1.4$\degree$, respectively.  The rotation axis is tilted out of the plane of the sky in the northern hemisphere. These values and orientation are nearly identical to the estimates found by the 2010 data sets.

\subsection{Comparing Results With Previous Work}
\label{subsec:litcomp}

Having demonstrated that starspot properties can be measured for $\lambda$ And using interferometric observations, these results are compared to the results of previous investigations. D95 created a surface map of $\lambda$ And via a matrix LCI technique using Johnson \emph{BV} light curves spanning one rotation period.  D95 models the observed light curve using 2 starspots with a \emph{T$_{R}$} = 0.83 $\pm$ 0.06.  One starspot is located at \emph{b} = 50$\degree$ with a $\phi$ = 8$\%$.  The other starspot is located at \emph{b} = 20$\degree$ with a $\phi$ = 4$\%$.  The starspots are separated by 140$\degree$ in longitude.  Both the latitudes and covering factors for these starspots are consistent with those identified in this work.  However, the temperature ratio in D95 is significantly less than that measured for both the 2010 (median \emph{T$_{R}$} = 0.961) and 2011 (median \emph{T$_{R}$} = 0.958) data sets.

A more recent study of the starspot properties of $\lambda$ And was performed by F08.  They use a 2-component LCI method using Johnson \emph{V}-Band photometry coupled with spectral line depth ratios to create a map of starspots on $\lambda$ And.  The results of F08 are very consistent with D95 with the modeled surface containing 2 cool starspots each with a \emph{T$_{R}$} = 0.815$^{0.064}_{-0.036}$.  The covering factors for the two starspots are 8.7$\%$ and 3.6$\%$ located at latitudes 57$\degree$ and 9$\degree$, respectively.  The starspots are separated by 81$\degree$ in longitude.  As with D95, the positions and covering factors of the starspots identified by F08 are consistent with the starspots imaged in this work.  However, the temperature ratio is significantly less.   One additional difference between F08 and this work, as well as D95, is the modeling of 2 plage regions by F08.  These bright regions are similar in size to the modeled cool starspots.  The plages are also in similar locations only offset to the starspots by $\sim$20$\degree$ in longitude and $\sim$7$\degree$ in latitude.  By design no hot starspot/plage regions were modeled in this work, however the reconstructed images show no convincing evidence these exist in any observed epoch.

\section{SUMMARY}
\label{sec:summary}

$\lambda$ Andromedae, a bright (V: 3.872 mag, H: 1.501 mag) G8 giant, has a long recorded history of consistent sinusoidal-like photometric variability.  This variability is believed to result from the rotational modulation of cool starspots.  Using light curve inversion techniques, the presence of starspots have been indirectly revealed \citep{donati95,frasca08}.  Long baseline optical/near infra-red interferometry has the potential to directly detect and measure the characteristics of these starspots.  LBI has already directly imaged a number of astrophysical systems (e.g. close binaries, circumstellar disks, rapidly rotating stellar surfaces) with unprecedented angular resolution.

$\lambda$ And was observed using the MIRC beam combiner on the CHARA Array for 26 days spanning from Aug 17$^{th}$, 2008 to Sep 24$^{th}$, 2011.  The observing strategy evolved over time due to upgrades in the MIRC beam combiner.  Contemporaneous photometric observations are also available from Sep 20$^{th}$, 2008 to Jan 20$^{th}$, 2011.  The photometry provides an independent relative estimate of starspot coverage that can be compared to the imaging results.

The CHARA array observations in 2010 employed an observing strategy designed to maximize the [u,v] coverage.  The images created from the 2010 data feature one to four starspots at each of the 6 epochs.  As an ensemble, the median value in the starspot covering factor is 7.6$\%$.  The median value of the temperature ratio between starspot and the surrounding photosphere is 0.853.  The starspot properties extracted from the reconstructed images are consistent with the modeled results to within the error bars.  The observing cadence between the 6 epochs in 2010 is between 6 to 9 days corresponding to 10.9$\%$ to 16.4$\%$ of the rotation period.  The observations span 71$\%$ of one rotation cycle.  By measuring the apparent motion of four starspots for 3 epochs, the rotation period is measured and the rotation axis is determined.  The rotation period based on starspot motion is 60 $\pm$ 13 days, which is consistent with the photometric rotation period of 54.8 $\pm$ 1.9 days.  The rotation axis is tilted out of the plane of the sky in the north with an inclination of 78.0 $\pm$ 1.5$\degree$ and a position angle of 20.0 $\pm$ 6.8$\degree$

A photometric \emph{V}-Band time series is available beginning 10 days after the last interferometric observation in 2010.  The short-cadence time series spans approximately two rotations of $\lambda$ And.  The flux variability in the modeled time series follows the behavior of the photometric time series quite well when the proper scaling factors are applied with the exception of a single epoch.  The discrepancy in this epoch is difficult to explain based on [u,v] coverage or data quality.

In 2011, $\lambda$ And was observed for a single night on 6 different nights using 6 telescopes simultaneously.  This strategy substantially increased the number of visibilities and closure phases obtained for each block of data compared to 2010.  However, the [u,v] coverage decreased by appropriately a factor of 2 since the data were only composed of a single night per epoch.  Consequently, the model and reconstructed images are not as consistent as those created from the 2010 data.  One to two starspots are identified in the model images for each epoch.  As an ensemble, the median value of the starspot coverage is 12$\%$ and the median temperature ratio is 0.842.  The 2011 observing cadence is 4 or 5 days corresponding to 7.3$\%$ and 9.2$\%$ of the rotation period spanning $\sim$40$\%$ of one rotation.  One starspot is imaged in all of the 6 epochs and is used to measure a rotation period of 54.0 $\pm$ 7.6 days, which is nearly identical to the photometric rotation period of 54.8 $\pm$ 1.9 days.  The rotation axis points out of the plane of the sky in the north with an inclination of 77.98 $\pm$ 0.18$\degree$ and a position angle of 23.0 $\pm$ 6.4$\degree$.  These values are in agreement with the orientation computed from the 2010 data set.

The photometric \emph{V}-Band time series starting in 2011 overlaps the last two interferometric epochs and spans $\sim$2.5 rotations of $\lambda$ And.  The flux variability in the modeled time series follows the behavior of the photometric time series only for half of the epochs.  The discrepant epochs can be made to agree with the photometry if the covering factor and/or the temperature ratios are changed for a single starspot.  The parameters are changed to be identical to parameters identified for the same starspot in a different epoch where the starspot is more centrally located.

Using the combined 2010 and 2011 data sets the measured angular diameter of $\lambda$ And is 2.759 $\pm$ 0.050 mas with a power law limb darkening coefficient of 0.229 $\pm$ 0.111.  At a Hipparcos trigonometric distance of 37.87 $\pm$ 0.21 mas, this yields a linear radius of 7.831$^{+0.067}_{-0.065}$ R$_{\sun}$.  The angular diameter is consistent with that previously found by \citet{nordgren99} (2.66 $\pm$ 0.08 mas).

The 2010 and 2011 data sets have provided a convincing picture of starspots on the surface of $\lambda$ And in support of the closure phase information and the variable light curve.  The agreement between the modeled and reconstructed starspot properties is typically within one error bar. In addition, the starspots produce a flux variability which is consistent with that observed photometrically just subsequent to the interferometric observations.  There is evidence to suggest that starspots imaged in one epoch are again imaged in subsequent epochs.  This provides starspot determined rotation periods consistent with the photometrically identified periods.  This method, additionally, allows for the complete characterization of the rotation axis of $\lambda$ And.  

The authors will now express our thanks to our anonymous referee for their conversation and comments.  We would also like to express our thanks to both Dr. Chris Farrington and P.J. Goldfinger for their smooth operation of the CHARA Array.  We would like to acknowledge funding from the National Science Foundation grants AST-0707927 and AST-1108963 awarded to the University of Michigan.  Additional NSF funding was provided through the AST­-1445935 and AST-1009643 grants awarded to Georgia State University. GWH acknowledges long-term support from NASA, NSF, Tennessee State University, and the State of Tennessee through its Centers of Excellence program.

%Main Figures

\clearpage

\clearpage

\clearpage

\clearpage
\begin{figure}
  \plotone{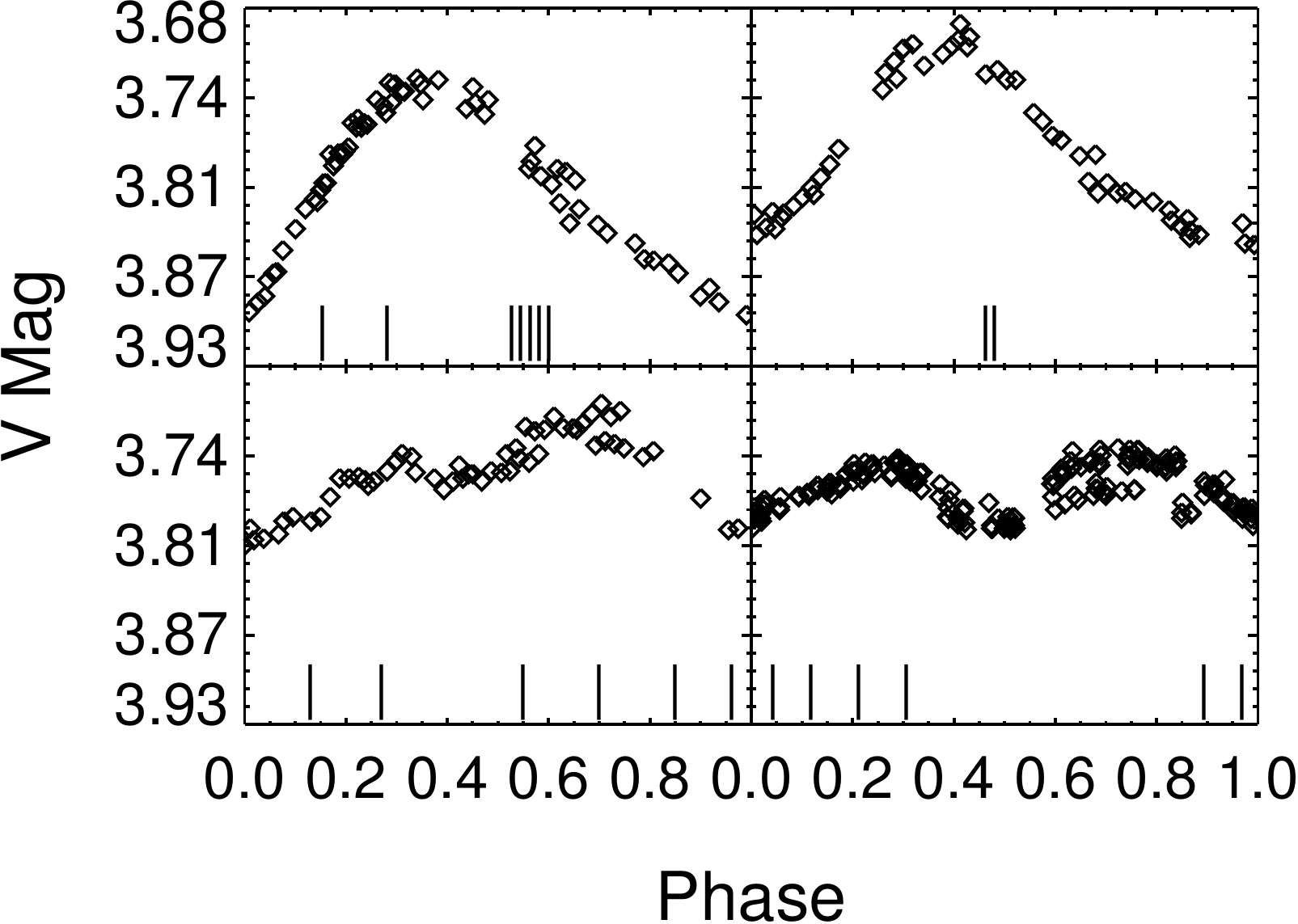}
  \caption{\emph{Top Left}: Season 2008 time series folded to a period of 54.27 $\pm$ 0.032 days.  \emph{Top Right}: Season 2009 time series folded to a period of 55.15 $\pm$ 0.91 days.  \emph{Bottom Left}: Season 2010 time series folded to a period of 53.4 $\pm$ 1.1 days.  \emph{Bottom Right}: Season 2011 time series folded to a period of 53.3 $\pm$ 1.9 days.}
  \label{fig:photper}
\end{figure}

\clearpage

\clearpage

\clearpage

\clearpage

\clearpage

\clearpage

\clearpage

\clearpage

\clearpage

\clearpage

\clearpage

\clearpage

\clearpage

\clearpage

\clearpage

\renewcommand{\thesection}{A-\arabic{section}}
\renewcommand{\thesubsection}{A-\arabic{section}.\arabic{subsection}}
\renewcommand{\thesubsubsection}{A-\arabic{subsubsection}}
\renewcommand{\thefigure}{A-\arabic{figure}}
% redefine the command that creates the figure and equation #s.
\setcounter{section}{0}
\setcounter{equation}{0}  % reset counter 
\setcounter{figure}{0} % reset figure number
\section*{APPENDIX. THE EARLY IMAGING ATTEMPTS}
\label{app:early_data}

This appendix discusses the failed attempts at high fidelity surface imaging to illustrate the effect of insufficient [u,v] coverage on interferometric starspot imaging.  Any successful program will need to be designed to maximize this coverage as much as possible.  The inconsistent results measured in 2008 are first discussed  and then followed by a discussion of the intriguing, yet inconclusive results measured in 2009.

\section{THE 2008 DATA SET}
\label{sec:2008_data}

\begin{figure}
  \plotone{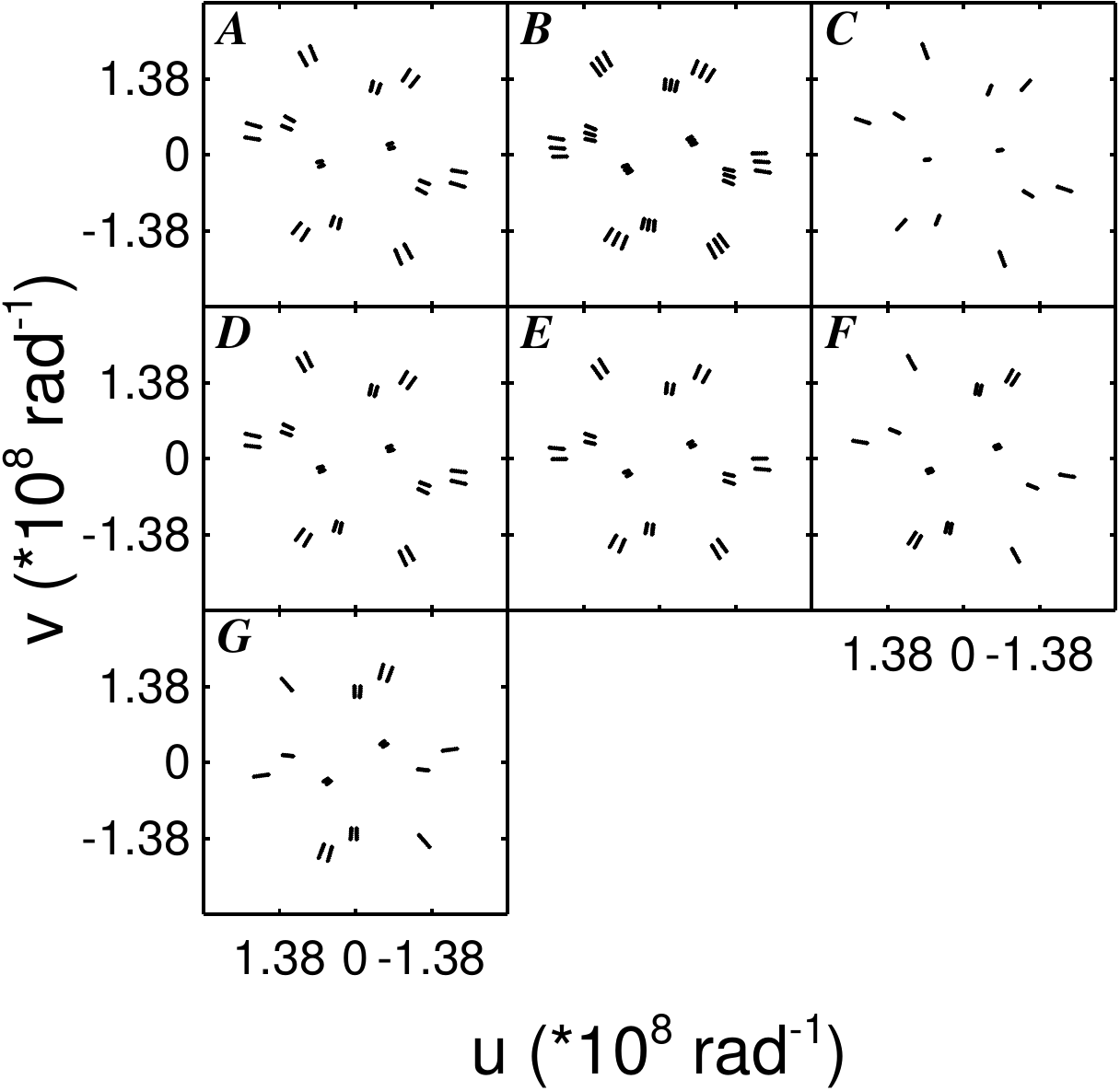}
  \caption{The [u,v] coverage obtained for the 2008 observing run.  A - Aug 17$^{th}$; \emph{2010}: B - Aug 18$^{th}$; C - Aug 19$^{th}$; D - Aug 20$^{th}$; E - Aug 21$^{st}$; F - Sep 20$^{th}$; G - Sep 27$^{th}$}
  \label{fig:uvplotone}
\end{figure}

In 2008, two observing runs of $\lambda$ And were performed with one in August and the other in September.  Both runs employed ``snapshot'' observations using the S1-E1-W1-W2 telescopes.  These observations consist of only two or three bracketed observations per night.  The August run was composed of observations taken on five consecutive nights between the 17$^{th}$ and the 21$^{st}$.  The [u,v] coverage achieved ranged from 48 to 144 data points with the densest coverage obtained on Aug 18$^{th}$.  Fig~\ref{fig:uvplotone} contains the plots of these [u,v] configurations.  The September run was composed of two observations taken a week apart on the 20$^{th}$ and the 27$^{th}$.  The [u,v] coverage achieved was 72 points for each night.  Fig~\ref{fig:uvplotone} contains the [u,v] configurations for these nights.  

\begin{figure}
  \plotone{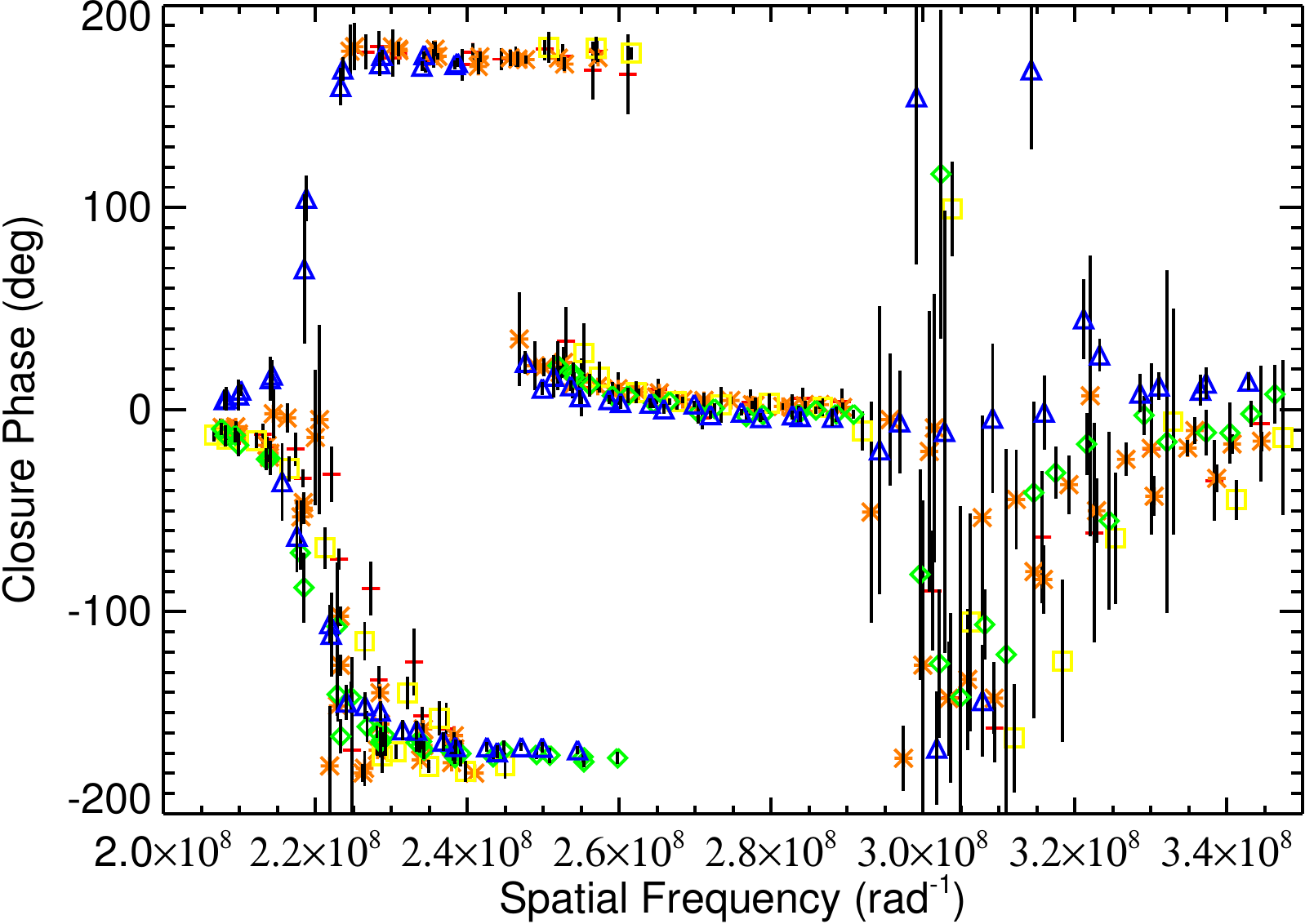}
  \caption{The observed closure phases for the Aug 2008 data sets.  \emph{Red Cross}: Aug 17$^{th}$. \emph{Orange Asterisks}: Aug 18$^{th}$. \emph{Yellow Squares}: Aug 19$^{th}$.  \emph{Green Diamonds}: Aug 20$^{th}$.  \emph{Blue Triangles}: Aug 21$^{st}$.  The distinct non-zero closure phase signature points to surface asymmetries.  The similarity in the closure phase between nights indicates a consistent asymmetric surface pattern from night to night.}
  \label{fig:oheightcp}
\end{figure}

Fig~\ref{fig:oheightcp} shows a distinct non-zero closure phase signature across most sampled spatial scales for all five days.  Unlike in 2010 and 2011, the closure phase signature changes only slightly from night to night, consistent with the hypothesis of surface asymmetries that have not evolved on short time scales.  An unspotted model image yields an extremely poor fit to the interferometric data for each epoch with the reduced $\chi^{2}$ ranging between 5.6 to 18.

\begin{figure}
  \plotone{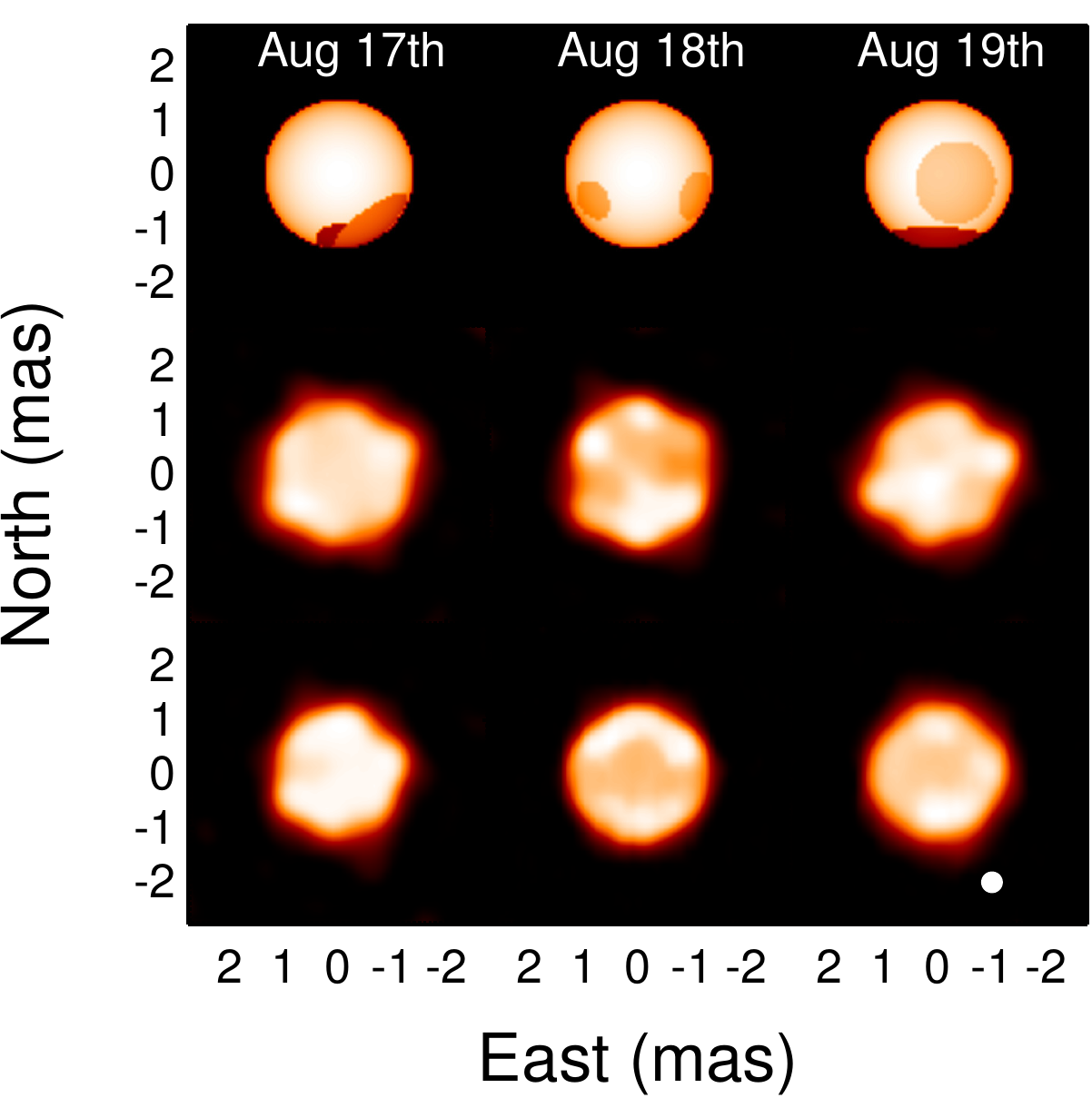}
  \caption{Results from the Aug 17$^{th}$, Aug 18$^{th}$, and Aug 19$^{th}$, 2008 data sets, including the model images (top row), reconstructed images (middle row), and simulated images (bottom row).  The white dot in the lower right hand corner represents the 0.4 mas resolution limit for the CHARA array.}
  \label{fig:oheightdataplota}
\end{figure}

The reduced $\chi^{2}$ for these models are all below 2.85 with the lowest reduced $\chi^{2}$ (1.14) occurring on Aug 19$^{th}$.  Despite this, these images do not present a consistent starspot configuration as can be seen in Fig~\ref{fig:oheightdataplota}.  This figure contains the model, reconstructed, and simulated images for Aug 17$^{th}$, 18$^{th}$, and 19$^{th}$.  The one day cadence should present nearly identical surface images as the surface only rotates 1.8$\%$ from night to night.  Starspot evolution on this time scale is not typical for magnetically active stars \citep{berdyugina05,strassmeier09}. Additionally, the reconstructed images do not even contain any conclusive evidence for starspots.    

The likely factors contributing to the non-detection of a consistent starspot configuration include, but are not limited to, poor [u,v] sampling, mis calibration and the influence of the binary 37 And used as a calibrator.

The phased photometric time series (Fig~\ref{fig:photper}) indicates these interferometric observations where taken $\sim$11 days after maximum brightness.  As shown in the study by \citet{henry95}, the maximum brightness of $\lambda$ And can vary by as much as 0.05 mag in the \emph{V}-Band.  This would indicate that even at time of maximum brightness the visible surface still contains starspots.  Therefore, the inconsistent images are probably not due to a insufficient starspot presence.

Thus, despite strong evidence for starspots on the surface of $\lambda$ And during these epochs, from both measured non-zero closure phases and the variable light curve, the [u,v] coverage using 4 telescopes on a single night is insufficient to confidently determine starspot properties. 

\section{THE 2009 DATA SET}
\label{sec:2009_data}

\begin{figure}
  \plotone{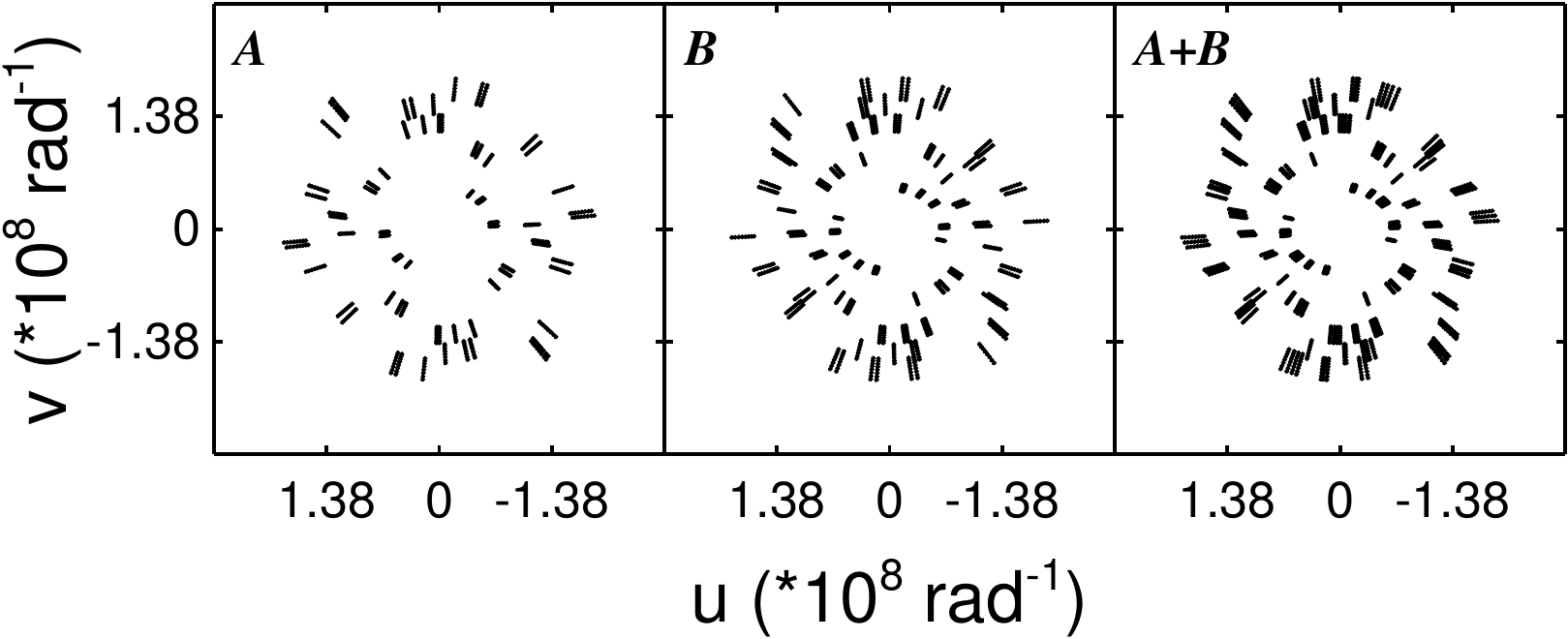}
  \caption{The [u,v] coverage obtained for the 2009 data sets.  A - Aug 24$^{nd}$; B - Aug 25$^{th}$}
  \label{fig:uvplotfive}
\end{figure}

The $\lambda$ And data set in 2009 consists of two observations on Aug 24$^{th}$ and Aug 25$^{th}$ that are combined to increase the final [u,v] coverage.  Given a rotation period of 55.15 days, starspots will migrate across the surface by $\sim$6$\degree$ over one night, so the combination of these two nights is not believed to adversely affect the quality of the extracted properties.  Each night is the combination of observations using both the S1-E1-W1-W2 and S2-E2-W1-W2 telescope arrays.  Fig~\ref{fig:uvplotfive} shows the distribution of the 704 [u,v] points obtained for the merged pair of observations.

Season 2009 spans 130.7 days or $\sim$2.4 rotation periods and has a $\Delta$V = 0.154 mag.  From one rotation to the next, the starspot properties do not appear to change significantly as illustrated by the low scatter compared to observation errors in Fig~\ref{fig:photper}.

\begin{figure}
  \plotone{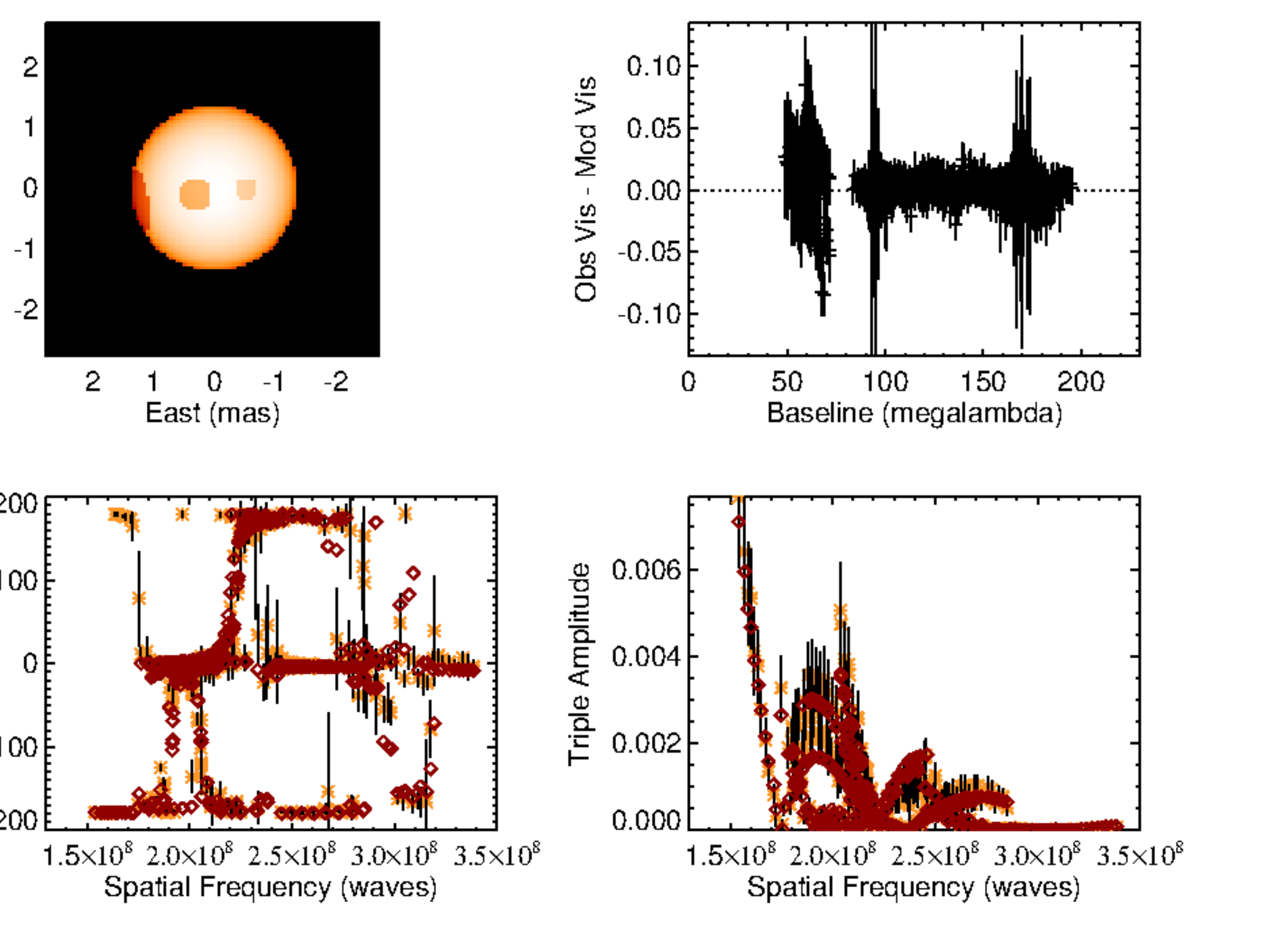}
  \caption{The best fit results for the Aug 24$^{th}$ + Aug 25$^{th}$, 2009 data sets.  \emph{Top Left}: The model image.  \emph{Top Right}: The observed minus modeled visibilities as a function of baseline.  \emph{Bottom Left}: The closure phase as a function of spatial frequency.  The orange asterisks indicate observed data and the red diamonds are the modeled fit. \emph{Bottom Right}: The triple amplitudes as a function of spatial frequency.  The symbols mean the same as in the closure phase plot.}
  \label{fig:ohninefitplot}
\end{figure}

Fig~\ref{fig:ohninefitplot} clearly show non-zero closure phases at both the lower and higher sampled spatial scales.  An unspotted model image does not fit well with the measured interferometric data resulting in a reduced $\chi^{2}$ = 5.9.

\begin{figure}
  \plotone{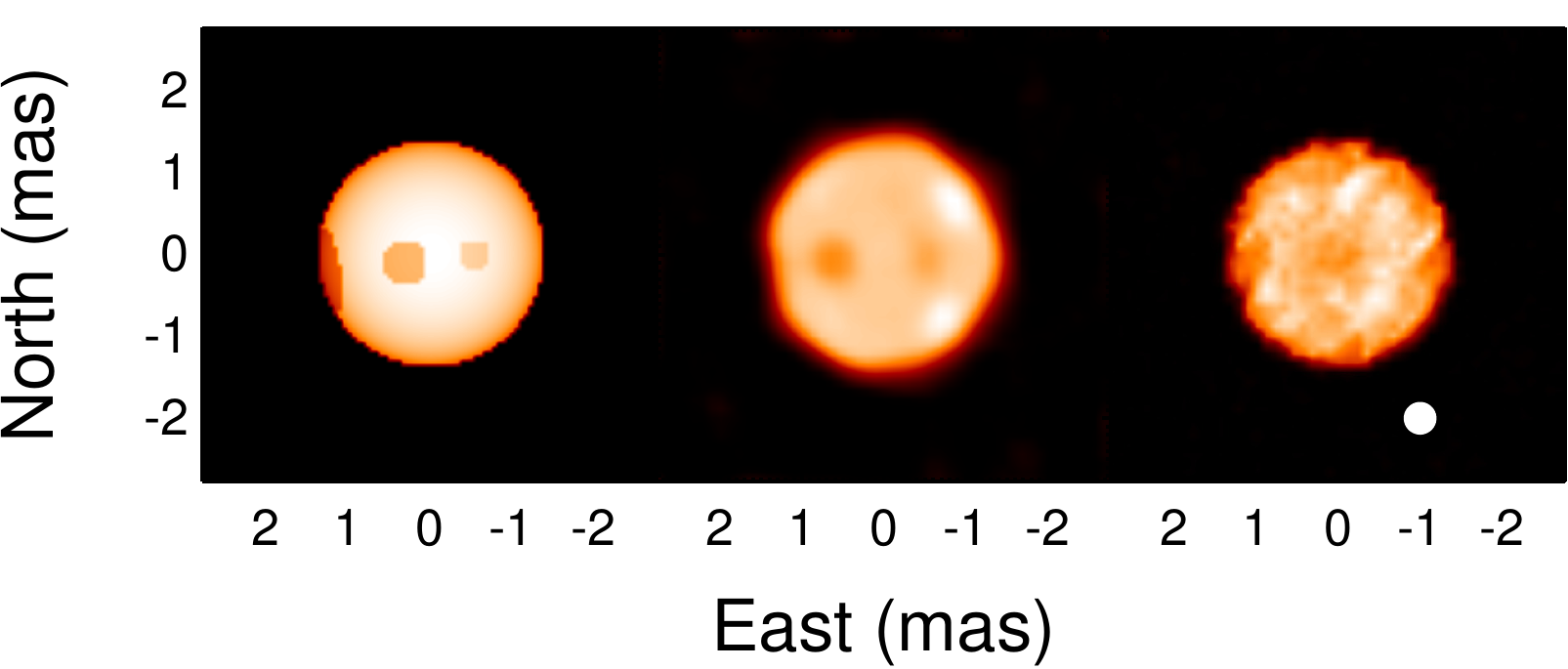}
  \caption{Results from the Aug 24$^{th}$ + Aug 25$^{th}$, 2009 data set, including the model images (left), reconstructed images (middle), and simulated images (right).  The white dot in the lower right hand corner represents the 0.4 mas resolution limit for the CHARA array.}
  \label{fig:ohninedataplot}
\end{figure}

The best-fit model (reduced $\chi^{2}$ = 1.44) contains three cool starspots.  Fig~\ref{fig:ohninefitplot} contains the best-fit model image along with the model fits to the visibilities, triple amplitudes and closure phases.  The starspot properties are listed in Table~\ref{tab:ohninedata}.  Fig~\ref{fig:ohninedataplot} contains the final model, reconstructed and simulated images for the 2009 data set.  The modeled starspot on the eastern limb was not conclusively detected in the reconstructed image.  The properties of the two reconstructed starspots are nearly identical to the corresponding modeled starspots, to within errors.  The western modeled and reconstructed starspots are included in the final results despite the measured covering factor of both being close to or below the CHARA array's angular resolution (0.4 mas or $\phi$ = 2.1$\%$).  The potential starspots are accepted as the model reduced $\chi^{2}$ is worse without its inclusion and the reconstructed starspot $\sigma_{ds}$ is 3.64.   

Two bright starspots are visible on the western limb; each starspot lies nearly equidistant above and below a east-west ``equator''.  These starspots are not visible in the simulated image.  This allows for the possibility these starspots are genuine surface features.   

Based on the phased time series (see Fig~\ref{fig:photper}), the interferometric observations were taken near maximum brightness.  However as discussed in $\S$~\ref{sec:2008_data}, it is possible for starspots to be present on the visible surface even at maximum brightness.  Given a rotation east to west, the identification of a large, cool starspot on the eastern limb would cause a drop in observed flux, as indicated by the photometry, as it rotates into view.

The observing strategy of combining two consecutive nights of data provides a near 500$\%$ gain in the best [u,v] coverage obtained in 2008.  This has produced a much improved consistency between the model image and the reconstructed image along with a better quality of fit in both cases.  However, the lack of multiple epochs does not provide consistency checks or a measure of the rotation period.

%Appendix Figures

\clearpage

\clearpage

\clearpage

\clearpage

\clearpage

\clearpage

%Tables

{\renewcommand{\arraystretch}{1.25}

\clearpage
\begin{deluxetable}{c c c l}
    \tablecolumns{4}
    \tablewidth{0pc}
    \tablecaption{CHARA Observing Log \label{tab:obs_log}}
    \tabletypesize{\scriptsize}
    \tablehead{
      \colhead{Date\tablenotemark{a}} & \colhead{Baselines} & \colhead{$\#$ of [u,v] points} & \colhead{Calibrators\tablenotemark{b}}
    }
    \startdata
      \emph{2008}&&&\\
      Aug 17$^{th}$ (54695.5)&S1-E1-W1-W2&96&37 And (2), 45 Per (3)\\
      Aug 18$^{th}$ (54696.5)&S1-E1-W1-W2&144&$\gamma$ Lyr, 7 And (2), 37 And, $\zeta$ Per (2)\\
      Aug 19$^{th}$ (54697.5)&S1-E1-W1-W2&48&7 And, $\zeta$ Per (2)\\
      Aug 20$^{th}$ (54698.5)&S1-E1-W1-W2&96&7 And (2), 37 And (2), 45 Per (3)\\
      Aug 21$^{st}$ (54699.5)&S1-E1-W1-W2&96&7 And (2), 37 And (2), $\zeta$ Per, 45 Per\\
      Sep 20$^{th}$ (54729.5)&S1-E1-W1-W2&72&7 And (2), $\zeta$ Cas, $\delta$ Per (2)\\
      Sep 27$^{th}$ (54736.5)&S1-E1-W1-W2&72&$\sigma$ Cyg, 37 And (2), $\zeta$ Per (2), tet Gem (3)\\
      \hline\\
      \emph{2009}&&&\\
      Aug 24$^{th}$ (55067.5)&S1-E1-W1-W2&272&7 And (3), 37 And (2)\\
      &S2-E2-W1-W2&&\\
      Aug 25$^{th}$ (55068.5)&S1-E1-W1-W2&432&7 And (4), 37 And (2), HR 75\\
      &S2-E2-W1-W2&&\\
      \hline\\
      \emph{2010}&&&\\
      Aug 2$^{nd}$ (55410.5)&S1-E1-W1-W2&168&7 And (2), 37 And\\
      &S2-E2-W1-W2&&\\
      Aug 3$^{rd}$ (55411.5)&S1-E1-W1-W2&456&$\sigma$ Cyg, 7 And (3), 37 And (2)\\
      &S2-E2-W1-W2&&\\
      Aug 10$^{th}$ (55418.5)&S1-E1-W1-W2&432&$\sigma$ Cyg, 7 And (3), 37 And (4)\\
      &S2-E2-W1-W2&&\\
      Aug 11$^{th}$ (55419.5)&S1-E1-W1-W2&288&$\sigma$ Cyg, 7 And (4), 37 And (2)\\
      &S2-E2-W1-W2&&\\
      Aug 18$^{th}$ (55426.5)&S1-E1-W1-W2&432&$\sigma$ Cyg, 7 And (3), 37 And (5)\\
      &S2-E2-W1-W2&&\\
      Aug 19$^{th}$ (55427.5)&S1-E1-W1-W2&432&$\sigma$ Cyg (2), 7 And (5), 37 And (7)\\
      &S2-E2-W1-W2&&\\
      Aug 24$^{th}$ (55432.5)&S1-E1-W1-W2&528&$\sigma$ Cyg (2), 7 And (6), 37 And (6)\\
      &S2-E2-W1-W2&&\\
      Aug 25$^{th}$ (55433.5)&S1-E1-W1-W2&384&$\sigma$ Cyg (2), 7 And (5), 37 And (2)\\
      &S2-E2-W1-W2&&\\
      Sep 2$^{nd}$ (55441.5)&S1-E1-W1-W2&528&7 And (7), 37 And (6)\\
      &S2-E2-W1-W2&&\\
      Sep 3$^{rd}$ (55442.5)&S1-E1-W1-W2&600&7 And (9), 37 And (3)\\
      &S2-E2-W1-W2&&\\
      Sep 10$^{th}$ (55449.5)&S1-E1-W1-W2&336&7 And (6), 37 And (2)\\
      &S2-E2-W1-W2&&\\
      \hline\\
      \emph{2011}&&&\\
      Sep 2$^{nd}$ (55806.5)&S1-S2-E1-E2-W1-W2&360&$\sigma$ Cyg, 7 And (2), 22 And (3), HR 653\\
      Sep 6$^{th}$ (55810.5)&S1-S2-E1-E2-W1-W2&392&$\sigma$ Cyg, 7 And (2), 22 And (3), HR 653\\
      Sep 10$^{th}$ (55814.5)&S1-S2-E1-E2-W1-W2&360&7 And (2), 22 And\\
      Sep 14$^{th}$ (55818.5)&S1-S2-E1-E2-W1-W2&864&7 And (4), 22 And\\
      Sep 19$^{th}$ (55823.5)&S1-S2-E1-E2-W1-W2&808&7 And (3), 22 And, HR 653 (2)\\
      Sep 24$^{th}$ (55828.5)&S1-S2-E1-E2-W1-W2&200&7 And, 22 And, HR 653 (2), $\eta$ Aur
    \enddata
    \tablenotetext{a}{The number in parenthesis is the approximate MJD}
    \tablenotetext{b}{The number in parenthesis is the number of observations during the night}
\end{deluxetable}

\begin{deluxetable}{c c c}

  \tablecolumns{3}
  \tablewidth{0pc}
  \tablecaption{Calibrator Angular Diameters \label{tab:cal_diam}}
  \tabletypesize{\scriptsize}
  \tablehead{
  \colhead{Calibrator} & \colhead{$\theta_{UD}$ (mas)} & \colhead{Reference}
  }

  \startdata
  37 And&0.682$\pm$0.03&new CHARA/MIRC measurement\\
  45 Per&0.41$\pm$0.02&\citep{barnes78}\\
  $\gamma$ Lyr&0.740$\pm$0.100&new CHARA/MIRC measurement\\
  7 And&0.663$\pm$0.024&new CHARA/MIRC measurement\\
  $\zeta$ Per&0.703$\pm$0.021&new CHARA/MIRC measurement\\
  $\zeta$ Cas&0.280$\pm$0.020&new CHARA/MIRC measurement\\
  $\delta$ Per&0.555$\pm$0.038&new CHARA/MIRC measurement\\
  $\sigma$ Cyg&0.542$\pm$0.021&new CHARA/MIRC measurement\\
  tet Gem&0.820$\pm$0.030&new CHARA/MIRC measurement\\
  HR 75&1.04$\pm$0.012&CHARM2 \citep{richichi05}\\
  22 And&0.591$\pm$0.041&SearchCal \citep{bonneau06}\\
  HR 653&0.646$\pm$0.045&SearchCal \citep{bonneau06}\\
  $\eta$ Aur&0.419$\pm$0.063&new CHARA/MIRC measurement
  \enddata

\end{deluxetable}

\begin{deluxetable}{c c c c c c c}

  \tablecolumns{7}
  \tablewidth{0pc}
  \tablecaption{Photometry Observing Log \label{tab:phot_log}}
  \tabletypesize{\scriptsize}
  \tablehead{
  \colhead{Season} & \colhead{Dates} & \colhead{T$_{0}$} & \colhead{$\Delta$T} & \colhead{N$_{obs}$} & \colhead{Period} & \colhead{$\Delta$\emph{V}}\\
  \colhead{} & \colhead{(MJD)} & \colhead{(MJD)} & \colhead{(days)} & \colhead{} & \colhead{(days)} & \colhead{(mag)}
  }

  \startdata
  2008&54729.7979 - 54861.6223&54722.0&131.8&66&54.27$\pm$0.032&0.165\\
  2009&55091.9383 - 55222.6037&55098.0&130.7&58&55.15$\pm$0.91&0.154\\
  2010&55459.7624 - 55581.6084&55488.7&121.8&73&53.35$\pm$1.1&0.099\\
  2011&55823.6466 - 55958.5967&55813.0&134.9&194&53.30$\pm$1.9&0.057
    \enddata

\end{deluxetable}

{\renewcommand{\arraystretch}{1}

\begin{deluxetable}{l c c c c c c}
    \tablecolumns{7}
    \tablewidth{0pc}
    \tablecaption{2010 Starspot Properties \label{tab:tendata}}
    \tabletypesize{\scriptsize}
    \tablehead{
      \multicolumn{7}{c}{MODEL}\\
      \hline\\
      \colhead{Param.} & \colhead{Epoch 1} & \colhead{Epoch 2} & \colhead{Epoch 3} & \colhead{Epoch 4} & \colhead{Epoch 5} & \colhead{Epoch 6}      
    }
    \startdata
    $\phi_{1}$ ($\%$)&5.3$\pm$5.3&7.6$\pm$4.0&8.4$\pm$5.0&7.9$\pm$3.7&13.8$\pm$4.6&44$\pm$12\\
    \emph{b$_{1}$} ($\degree$)&-10.1$\pm$1.3&0.5$\pm$1.3&7.1$\pm$1.2&11.5$\pm$1.1&4.9$\pm$1.017&55$\pm$0.98\\
    \emph{l$_{1}$} ($\degree$)&-55.12$\pm$0.84&-59.4$\pm$1.0&-64.9$\pm$5.4&-27.17$\pm$0.79&-46.8$\pm$1.8&-40.9$\pm$5.6\\
    \emph{T$_{R1}$}&0.926$\pm$0.013&0.758$\pm$0.018&0.768$\pm$0.016&0.772$\pm$0.036&0.870$\pm$0.059&0.925$\pm$0.017\\
    \emph{Starspot}&\emph{C}&\emph{E}&\emph{F}&\emph{F}&\emph{G}&---\\
    \hline\\
    $\phi_{2}$ ($\%$)&20.5$\pm$3.8&5.8$\pm$3.1&7.4$\pm$5.1&7.1$\pm$4.6&8.0$\pm$8.0&13.5$\pm$6.4\\
    \emph{b$_{2}$} ($\degree$)&23.5$\pm$1.3&-1.1$\pm$1.1&16.4$\pm$1.1&34.9$\pm$1.4&31.4$\pm$1.4&29.84$\pm$0.96\\
    \emph{l$_{2}$} ($\degree$)&3.12$\pm$0.75&-19.0$\pm$1.6&-16.4$\pm$1.7&19.00$\pm$0.96&24.0$\pm$1.4&9.7$\pm$4.9\\
    \emph{T$_{R2}$}&0.893$\pm$0.007&0.860$\pm$0.015&0.758$\pm$0.019&0.756$\pm$0.057&0.790$\pm$0.059&0.898$\pm$0.048\\
    \emph{Starspot}&\emph{B}&\emph{D}&\emph{E}&\emph{E}&\emph{F}&\emph{G}\\
    \hline\\
    $\phi_{3}$ ($\%$)&11.5$\pm$5.5&5.0$\pm$2.6&5.3$\pm$3.4&6.3$\pm$4.4&---&---\\
    \emph{b$_{3}$} ($\degree$)&52.5$\pm$2.1&-6.8$\pm$1.0&11.0$\pm$1.0&28.3$\pm$1.5&---&---\\
    \emph{l$_{3}$} ($\degree$)&76.4$\pm$6.9&10.6$\pm$1.2&30.4$\pm$1.7&70.8$\pm$7.8&---&---\\
    \emph{T$_{R3}$}&0.801$\pm$0.042&0.860$\pm$0.015&0.853$\pm$0.014&0.853$\pm$0.0105&---&---\\
    \emph{Starspot}&\emph{A}&\emph{C}&\emph{D}&\emph{D}&---&---\\
    \hline\\
    $\phi_{4}$ ($\%$)&---&21.8$\pm$5.8&4.0$\pm$4.1&---&---&---\\
    \emph{b$_{4}$} ($\degree$)&---&54.95$\pm$0.99&11.8$\pm$2.8&---&---&---\\
    \emph{l$_{4}$} ($\degree$)&---&77.5$\pm$1.5&68.1$\pm$6.7&---&---&---\\
    \emph{T$_{R4}$}&---&0.907$\pm$0.011&0.853$\pm$0.017&---&---&---\\
    \emph{Starspot}&---&\emph{B}&\emph{C}&---&---&---\\
    \hline\\
    $\chi^{2}_{red}$&4.92&2.87&2.88&4.14&5.35&4.79\\
    \hline\\
    \multicolumn{7}{c}{SQUEEZE}\\
    \hline\\
    $\phi_{1}$ ($\%$)&---&---&---&4.8 (-3.1)&5.8 (-8.0)&---\\
    \emph{b$_{1}$} ($\degree$)&---&---&---&3.4 (-8.1)&4.6 (-0.3)&---\\
    \emph{l$_{1}$} ($\degree$)&---&---&---&-28.74 (-1.57)&-34.2 (12.6)&---\\
    \emph{T$_{R1}$}&---&---&---&0.724 (-0.048)&0.817 (-0.053)&---\\
    \emph{$\sigma_{ds}$}&---&---&---&11.55&8.35&---\\
    \emph{Starspot}&---&---&---&\emph{F}&\emph{G}&---\\
    \hline\\
    $\phi_{2}$ ($\%$)&9.0 (-11.5)&4.0 (-1.8)&5.76 (-1.64)&4.8 (-2.3)&5.8 (-2.2)&6.8 (-6.7)\\
    \emph{b$_{2}$} ($\degree$)&23.6 (-0.1)&-2.3 (-1.2)&16.3 (-0.1)&28.7 (-6.2)&27.4 (-4.0)&21.10 (-8.74)\\
    \emph{l$_{2}$} ($\degree$)&2.5 (-0.62)&-27.4 (-8.4)&-22.0 (-5.6)&7.86 (-11.14)&11.7 (-12.3)&0.0 (-9.7)\\
    \emph{T$_{R2}$}&0.856 (-0.037)&0.856 (-0.004)&0.921 (0.163)&0.724 (-0.032)&0.761 (-0.029)&0.849 (-0.049)\\
    $\sigma_{ds}$&4.12&7.11&16.20&12.17&9.81&5.46\\
    \emph{Starspot}&\emph{B}&\emph{D}&\emph{E}&\emph{E}&\emph{F}&\emph{G}\\
    \hline\\
    $\phi_{3}$ ($\%$)&---&4.0 (-1.0)&3.6 (-1.7)&---&---&---\\
    \emph{b$_{3}$} ($\degree$)&---&-11.5 (-4.7)&4.6 (-6.4)&---&---&---\\
    \emph{l$_{3}$} ($\degree$)&---&9.4 (-1.2)&28.8 (-1.6)&---&---&---\\
    \emph{T$_{R3}$}&---&0.856 (-0.004)&0.849 (-0.004)&---&---&---\\
    $\sigma_{ds}$&---&6.28&8.90&---&---&---\\
    \emph{Starspot}&---&\emph{C}&\emph{D}&---&---&---\\
    \hline\\
    $\phi_{4}$ ($\%$)&---&---&2.6 (-1.4)&---&---&---\\
    \emph{b$_{4}$} ($\degree$)&---&---&-30.0 (-41.8)&---&---&---\\
    \emph{l$_{4}$} ($\degree$)&---&---&2.7 (-65.4)&---&---&---\\
    \emph{T$_{R4}$}&---&---&0.915 (0.062)&---&---&---\\
    $\sigma_{ds}$&---&---&6.16&---&---&---\\
    \emph{Starspot}&---&---&\emph{C}&---&---&---\\  
    \hline\\
    $\chi^{2}_{red}$&1.00&1.00&1.03&0.98&0.97&0.99
    \enddata
    \tablecomments{The parenthetical number beside the SQUEEZE parameter values is the difference between the SQUEEZE and model values for the same starspot in each image.  For $\phi_{2}$ and \emph{T$_{R}$} values, a positive number indicates a larger value for the SQUEEZE parameter.  For \emph{b} and \emph{l}, a positive number indicates the starspot in the SQUEEZE image is further north and west, respectively.}
    
 \end{deluxetable}

} 
 
\begin{deluxetable}{c c c c c}
    \tablecolumns{5}
    \tablewidth{0pc}
    \tablecaption{Evidence of Stellar Rotation in the 2010 Data Set \label{tab:rotation10}}
    \tabletypesize{\scriptsize}
    \tablehead{
      \colhead{Starspot} & \colhead{Epoch Range} & \colhead{$\Delta\phi$} & \colhead{$\Delta$T$_{R}$} & \colhead{P$_{rot}$} \\
      \colhead{} & \colhead{} & \colhead{\emph{($\%$)}} & \colhead{} & \colhead{\emph{(days)}}
    }
    \startdata
      A&1$\rightarrow$?&---&---&---\\
      \hline\\
      B&1$\rightarrow$2&1&-0.014&46.6\\
      \hline\\
      C&1$\rightarrow$2&-0.3&-0.066&44.3\\
      &2$\rightarrow$3&-1&-0.007&47.9\\
      \hline\\
      D&2$\rightarrow$3&-0.5&-0.007&56.9\\
      &3$\rightarrow$4&1&0.000&70.1\\
      \hline\\
      E&2$\rightarrow$3&-0.2&0.000&63.7\\
      &3$\rightarrow$4&-0.3&-0.002&78.7\\
      \hline\\
      F&3$\rightarrow$4&-0.5&0.014&77.0\\
      &4$\rightarrow$5&0.1&0.018&63.4\\
      \hline\\
      G&5$\rightarrow$6&0&0.028&49.1
    \enddata
\end{deluxetable}

\begin{deluxetable}{l c c c c c c}
    \tablecolumns{7}
    \tablewidth{0pc}
    \tablecaption{2011 Starspot Properties \label{tab:elevendata}}
    \tabletypesize{\scriptsize}
    \tablehead{
      \multicolumn{7}{c}{MODEL}\\
      \hline\\
      \colhead{Param.} & \colhead{Epoch 1} & \colhead{Epoch 2} & \colhead{Epoch 3} & \colhead{Epoch 4} & \colhead{Epoch 5} & \colhead{Epoch 6}      
    }
    \startdata
    $\phi_{1}$ ($\%$)&16.9$\pm$5.7&12.5$\pm$4.8&11.8$\pm$3.1&10.2$\pm$2.5&10.4$\pm$6.4&14.5$\pm$3.8\\
    \emph{b$_{1}$} ($\degree$)&-1.3$\pm$2.1&4.15$\pm$0.93&11.9$\pm$1.3&24.9$\pm$0.6&1.30$\pm$0.92&14.6$\pm$1.6\\
    \emph{l$_{1}$} ($\degree$)&-60.3$\pm$5.4&-35.6$\pm$1.6&-7.4$\pm$1.0&11.8$\pm$1.3&-60.7$\pm$3.5&-34.1$\pm$1.2\\
    \emph{T$_{R1}$}&0.825$\pm$0.014&0.825$\pm$0.015&0.797$\pm$0.013&0.860$\pm$0.007&0.825$\pm$0.011&0.825$\pm$0.015\\
    \emph{Starspot}&\emph{B}&\emph{B}&\emph{B}&\emph{B}&\emph{C}&\emph{C}\\
    \hline\\
    $\phi_{2}$ ($\%$)&10.0$\pm$4.5&10.0$\pm$5.4&14.7$\pm$5.7&---&14.8$\pm$5.3&15.6$\pm$6.0\\
    \emph{b$_{2}$} ($\degree$)&3.65$\pm$0.95&20.1$\pm$1.7&37.7$\pm$1.4&---&40.5$\pm$1.4&62.1$\pm$2.6\\
    \emph{l$_{2}$} ($\degree$)&8.58$\pm$0.76&38.1$\pm$1.2&68.1$\pm$2.2&---&49.8$\pm$2.9&85.6$\pm$5.7\\
    \emph{T$_{R2}$}&0.864$\pm$0.013&0.864$\pm$0.015&0.867$\pm$0.015&---&0.842$\pm$0.014&0.849$\pm$0.016\\
    \emph{Starspot}&\emph{A}&\emph{A}&\emph{A}&---&\emph{B}&\emph{B}\\
    \hline\\
    $\chi^{2}_{red}$&2.91&3.29&3.81&4.13&6.24&5.69\\
    \hline\\
    \multicolumn{7}{c}{SQUEEZE}\\
    \hline\\
    $\phi_{1}$ ($\%$)&---&4.0 (-8.5)&7.8 (-4.0)&4.8 (-5.4)&4.0 (-6.4)&---\\
    \emph{b$_{1}$} ($\degree$)&---&11.54 (7.39)&16.3 (4.4)&24.8 (-0.1)&-22.33 (-23.63)&---\\
    \emph{l$_{1}$} ($\degree$)&---&-29.3 (6.3)&-14.5 (-7.1)&6.3 (-5.5)&14.4 (75.1)&---\\
    \emph{T$_{R1}$}&---&0.839 (0.014)&0.751 (-0.046)&0.701 (-0.159)&0.849 (0.024)&---\\
    $\sigma_{ds}$&---&9.09&5.59&9.67&3.23&---\\
    \emph{Starspot}&---&\emph{B}&\emph{B}&\emph{B}&\emph{C}&---\\
    \hline\\
    $\phi_{2}$ ($\%$)&4.8 (-5.2)&---&---&2.6&4.8 (-10.0)&---\\
    \emph{b$_{2}$} ($\degree$)&9.21 (5.56)&---&---&3.4&24.8 (-15.7)&---\\
    \emph{l$_{2}$} ($\degree$)&12.88 (4.30)&---&---&-31.4&5.1 (-44.7)&---\\
    \emph{T$_{R2}$}&0.860 (-0.004)&---&---&0.915&0.849 (0.007)&---\\
    $\sigma_{ds}$&5.31&---&---&3.38&3.18&---\\
    \emph{Starspot}&\emph{A}&---&---&\emph{B}&\emph{B}&---\\
    \hline\\
    $\chi^{2}_{red}$&1.01&1.02&0.98&0.99&0.95&0.97
    \enddata
    \tablecomments{The parenthetical number beside the SQUEEZE parameter values is the difference between the SQUEEZE and model values for the same starspot in each image.  For $\phi_{2}$ and \emph{T$_{R}$} values, a positive number indicates a larger value for the SQUEEZE parameter.  For \emph{b} and \emph{l}, a positive number indicates the starspot in the SQUEEZE image is further north and west, respectively.}

 \end{deluxetable}

\begin{deluxetable}{c c c c c}
    \tablecolumns{5}
    \tablewidth{0pc}
    \tablecaption{Evidence of Stellar Rotation in the 2011 Data Set \label{tab:rotation11}}
    \tabletypesize{\scriptsize}
    \tablehead{
      \colhead{Starspot} & \colhead{Epoch Range} & \colhead{$\Delta\phi$} & \colhead{$\Delta$T$_{R}$} & \colhead{P$_{rot}$} \\
      \colhead{} & \colhead{} & \colhead{\emph{($\%$)}} & \colhead{} & \colhead{\emph{(days)}}
    }
    \startdata
      A&1$\rightarrow$2&0&0.000&43.4\\
      &2$\rightarrow$3&5&0.003&45.9\\
      \hline\\
      B&1$\rightarrow$2&-4&0.000&56.7\\
      &2$\rightarrow$3&-1&-0.028&49.8\\
      &3$\rightarrow$4&-2&0.063&64.6\\
      &4$\rightarrow$5&5&-0.035&51.1\\
      &5$\rightarrow$6&1&0.007&59.2\\
      \hline\\
      C&5$\rightarrow$6&5&0.000&61.2
   \enddata
\end{deluxetable}

\setcounter{table}{0} % reset figure number
\renewcommand{\thetable}{A1-\arabic{table}}

\begin{deluxetable}{c c c c}
    \tablecolumns{4}
    \tablewidth{0pc}
    \tablecaption{2009 Starspot Properties \label{tab:ohninedata}}    
    \tabletypesize{\scriptsize}
    \tablehead{
      \colhead{} & \colhead{} & \multicolumn{2}{c}{Aug 24$^{th}$+25$^{th}$} \\
      \colhead{Parameter} &\colhead{} & \colhead{Model} & \colhead{SQUEEZE}
    }
    \startdata
      $\phi_{1}$&($\%$)&16.0$\pm$1.9&---\\
      \emph{b$_{1}$} &(\degree)&-8.6$\pm$2.0&---\\
      \emph{l$_{1}$}&(\degree)&-76.0$\pm$1.9&---\\
      \emph{T$_{R1}$}&&0.961$\pm$0.008&---\\
      $\sigma$&&--- &---\\
      \hline\\
      $\phi_{2}$&($\%$)&4.1$\pm$6.8&3.6\\
      \emph{b$_{2}$}&(\degree)&-2.3$\pm$1.9&-3.4\\
      \emph{l$_{2}$}&(\degree)&-13.4$\pm$5.3&-26.8\\
      \emph{T$_{R2}$}&&0.931$\pm$0.024&0.922\\
      $\sigma$&&--- &4.8\\
      \hline\\
      $\phi_{3}$&($\%$)&2.2$\pm$2.2&2.0\\
      \emph{b$_{3}$}&(\degree)&-0.8$\pm$1.4&-2.3\\
      \emph{l$_{3}$}&(\degree)&22.9$\pm$2.0&21.1\\
      \emph{T$_{R3}$}&&0.979$\pm$0.010&0.924\\
      $\sigma$&&---&3.6\\
      \hline\\
      Reduced $\chi^{2}$&&1.44&0.93
    \enddata
\end{deluxetable}

}

\bibliography{lam_and}

\end{document}